\pdfoutput=1

\documentclass[useAMS,usenatbib,fleqn]{mnras}

\usepackage{graphicx,amssymb}
\usepackage{epsf,psfig}
\usepackage{natbib}
\usepackage{amsmath}

\title[The role of NHIMs in barred galaxies]{Orbital and escape dynamics in barred galaxies - II. \\ The 3D system: Exploring the role of the normally \\ hyperbolic invariant manifolds}

\author[Ch. Jung \& E. E. Zotos]{Christof Jung$^1$\thanks{E-mail:jung@fis.unam.mx} and Euaggelos E. Zotos$^2$\thanks{E-mail: evzotos@physics.auth.gr} \\
$^1$ Instituto de Ciencias F\'{i}sicas, Universidad Nacional Aut\'{o}noma de M\'{e}xico
Av. Universidad s/n, 62251 Cuernavaca, Mexico \\
$^2$ Department of Physics, School of Science, Aristotle University of Thessaloniki,
GR-541 24, Thessaloniki, Greece
}

\begin{document}

\date{Accepted 2016 September 7. Received 2016 August 22; in original form 2016 April 21}

\pubyear{2016} \volume{463} \pagerange{3965--3988}

\setcounter{page}{3965}

\maketitle

\label{firstpage}

\begin{abstract}
A three degrees of freedom (3-dof) barred galaxy model composed of a spherically symmetric nucleus, a bar, a flat disc and a spherically symmetric dark matter halo is used for investigating the dynamics of the system. We use colour-coded plots to demonstrate how the value of the semi-major axis of the bar influences the regular or chaotic dynamics of the 3-dof system. For distinguishing between ordered and chaotic motion we use the Smaller ALingment Index (SALI) method, a fast yet very accurate tool. Undoubtedly, the most important elements of the dynamics are the normally hyperbolic invariant manifolds (NHIMs) located in the vicinity of the index 1 Lagrange points $L_2$ and $L_3$. These manifolds direct the flow of stars over the saddle points, while they also trigger the formation of rings and spirals. The dynamics in the neighbourhood of the saddle points is visualized by bifurcation diagrams of the Lyapunov orbits as well as by the restriction of the Poincar\'e map to the NHIMs. In addition, we reveal how the semi-major axis of the bar influences the structure of these manifolds which determine the final stellar structure (rings or spirals). Our numerical simulations suggest that in galaxies with weak bars the formation of $R_1$ rings or $R_1'$ pseudo-rings is favoured. In the case of galaxies with intermediate and strong bars the invariant manifolds seem to give rise to $R_1R_2$ rings and twin spiral formations, respectively. We also compare our numerical outcomes with earlier related work and with observational data.
\end{abstract}


\section{Introduction}
\label{intro}

Observations suggest that in the central regions of most disc and spiral galaxies we often encounter bars which are linear extended stellar structures. In the mid 60s de Vaucouleurs \citep{dV63} revealed that about one third of the observed disc galaxies do not contain a bar, one third have intermediate or undeterminable types of bars, while the remaining third exhibit strong bar properties \citep[e.g.,][]{Ee00,SRSS03}. Furthermore, recent observational data indicate that the fraction of spiral galaxies, which contain a bar in their central region, reduces with increasing redshift \citep[e.g.,][]{MNH11,MM13,SEE08}. The orbits of the stars in the interior region are usually disturbed by density waves radiating from the center of the galaxy. This phenomenon is believed to be the main dynamical reason responsible for the occurrence of bars in disc and spiral galaxies.

For many years the only realistic model for galactic bars was the Ferrers' triaxial model \citep{F77}. However, the corresponding potential derived through the density distribution is too complicated, while it is not known in a closed form. On this basis, in \citet{JZ15} we decided to introduce a new analytical model for describing galactic bars. Following the work of \citet{P84} we used a multi component model describing the motion of stars in a barred galaxy with a central spherically symmetric nucleus with an additional flat disc. Our new bar potential is much simpler with respect to Ferrers' however it is still very realistic, thus having a clear advantage on the performance speed of the numerical calculations.

In \citet{JZ16} (hereafter Paper I) we added a fourth component corresponding to a spherical dark matter halo for obtaining a realistic asymptotic behaviour for large galactocentric distances. In Paper I, which is the first part of the series, we explored the escape dynamics in the 2 degrees of freedom (2-dof) system. In particular, we conducted a thorough and systematic orbit classification in several types of planes. We managed to locate the basins of escape through the two symmetrical escape channels around the Lagrange points $L_2$ and $L_3$ and also to relate them to the corresponding distribution of the escape rates of the orbits, following the numerical methods used in \citet{EP14}. We also presented evidence that the unstable manifolds which guide the orbits in and out the interior region of the galaxy are directly related to the formation of spiral and ring stellar structures observed in real barred galaxies.

When we want to understand the dynamics of a Hamiltonian 2-dof system, then a standard procedure is to construct its Poincar\'e map for a fixed energy, which has a 2-dimensional domain. To produce the plot we select a moderate number of initial points and plot the iterated images of these initial points. In an integrable system the domain of the map has an invariant foliation into subsets of half the dimension. For a 2-dimensional map these invariant subsets are 1-dimensional curves. Accordingly the iterates of any initial point lie on a 1-dimensional curve. If integrability is destroyed, then also
this foliation is destroyed and at least for some initial conditions the iterates cover 2-dimensional subsets of the domain. The relative ratio of area covered by invariant lines gives an impression how close the system is to integrability and how
important are the perturbations away from integrability. In general the Poincar\'e map can not be constructed analytically even in most of the integrable cases and for Hamiltonians of simple functional structure. The only important exception appears for periodically delta kicked systems where it is trivial to write down the map in closed form.

Usually the most important elements of the dynamics of the map are hyperbolic fixed points, they are the most important elements of the skeleton of the dynamics. They have 1-dimensional stable and unstable manifolds which divide the domain of the map into regions of different behaviour. The intersection pattern between the stable and unstable manifolds traces out horseshoe constructions in the map and it is a good starting point for the construction of a symbolic dynamics. All these properties are explained well in standard books on dynamical systems like \citet{J91} and \citet{LL93}, while for pictorial explanations the reader is referred to \citet{AS92}. Essential properties of the hyperbolic fixed points to achieve this are: They are invariant subsets of codimension 2 in the domain of the map and they have stable and unstable manifolds of codimension 1.

When we go over to systems with more degrees of freedom then the question arises which objects in the higher dimensional Poincar\'e map take over the important role which hyperbolic fixed points play in 2-dimensional maps. In the present article we deal with the motion of test particles in a 3-dimensional position space under the influence of some effective potential, i.e. we are interested in 3-dof systems. Then the Poincar\'e map for a fixed energy acts on a 4-dimensional domain and in the following we will concentrate on this case. To direct and channel the general dynamics the higher dimensional generalizations of hyperbolic fixed points should have the same important properties which we have pointed out at the end of the previous paragraph for the hyperbolic fixed points in 2-dimensional maps. They should be invariant, should be of codimension 2 and should have stable and unstable manifolds of codimension 1. Accordingly, in our 4-dimensional map we must be looking for invariant 2-dimensional surfaces in the domain which are hyperbolic in the directions normal to the surface itself. Such surfaces are special examples of objects which in mathematics are known for a long time under the name Normally Hyperbolic Invariant Manifolds (usual abbreviation NHIMs). More details regarding the mathematical background of the NHIMs can be found in \citet{W94}. Also the 4-dimensional map has hyperbolic fixed points. But their stable and unstable manifolds are 2-dimensional. Therefore they do not divide anything in the 4-dimensional domain of the map. And therefore they are of little importance for the overall behaviour of the map.

The next natural question is: Where do we usually find such objects? The answer is: Usually they live over index 1 saddles of the effective potential, i.e. over saddles with a single coordinate along which the potential goes down (into both orientations of this coordinate). Here we see already that this situation is frequently given in systems of celestial mechanics when we treat them in a rotating coordinate system. Then these systems have Lagrange points $L_2$ and $L_3$ which usually are index 1 saddles. In the present article we treat a barred galaxy in the rotating frame and we will see that all by itself we run exactly into this situation. The main topic of this article will be, to show, how NHIMs appear over the Lagrange points $L_2$ and $L_3$ and how they have global influence on the dynamics. In the example of the barred galaxy the main result will be how the unstable manifolds of the NHIMs are related to the rings and spirals of the galaxy. So we see how such abstract mathematical objects have implications which are directly observed.

The invariant manifolds direct the flow over the saddle equilibrium points and they determine how general orbits enter the potential interior and leave it again. The invariant manifolds can be thought of as tubes which control the motion of test particles (stars in the case of a galaxy) with the same value of energy as the manifolds \citep[see e.g.,][]{GKL04,KLMR00}. The NHIM has an inner branch of the stable manifold, an inner branch of the unstable manifold, an outer branch of the stable manifold and an outer branch of the unstable manifold. The two outer branches are outside of corotation and the two inner branches are inside of corotation. The stable manifold consists of orbits which converge to the NHIM in the future and the unstable manifold consists of orbits which converge to the NHIM in the past. At this point, we would like to clarify that the terms stable and unstable manifolds do not mean, by no means, that the orbits of stars inside them are stable and unstable, respectively.

Usually in disc galaxies we observe interesting stellar structure. Undoubtedly, the most spectacular ones are the rings and the spirals. Observations strongly indicate that a large percentage of disc galaxies (about 70\%, according to recent measurements) exhibit bar-like formations \citep[e.g.,][]{Ee00,SRSS03}. In particular, bars possess two arms that very often start from the two ends of the galactic bar and then wind outwards thus covering a substantial region of the disc. Rings in barred galaxies on the other hand, are directly associated with the formation of new stars \citep[e.g.,][]{KBHSd95,MKVR08,Se10,HMLHOW11}. According to their geometry and size the rings are classified into three main categories: (i) small nuclear rings which surround the central nucleus, (ii) inner rings which are slightly elongated along the bar with comparable size to it and (iii) outer rings with a major axis with about twice the size of the bar. It was found that the stable and the unstable manifolds of the Lyapunov periodic orbits \citep{L07} are responsible for the formation of rings and spirals \citep[e.g.,][]{ARGM09,ARGBM09,ARGBM10,ARGM11,RGMA06,RGAM07}.

The present paper is organized as follows: In Section \ref{galmod} we briefly describe the main properties of the four component galaxy model. In the following Section, we investigate how the semi-major axis of the galactic bar influences the orbital properties of the 3-dof system. Section \ref{nhims} contains a detailed description of the dynamics in the
neighbourhood of the saddle point $L_2$ and in particular of the NHIM sitting near this Lagrange point. In Section \ref{rs} we link the invariant manifolds with the ring and the spiral structures observed in barred galaxies. Our paper ends with Section \ref{conc}, where the main conclusions of our work are presented. An appendix contains, for the nonspecialists, the explanations of some concepts and terms from dynamical system theory which we use frequently during the article and which are essential for the understanding of our work.

\section{Presentation of the galactic model}
\label{galmod}

Let us briefly recall the multi-component model for the description of the motion of stars in barred galaxies which was introduced in \citet{JZ15} and upgraded in Paper I. The four components of the total gravitational potential, $\Phi_{\rm t}(x,y,z)$, are the following:

\begin{itemize}
  \item A spherically symmetric nucleus described by a Plummer potential \citep{BT08}
  \begin{equation}
  \Phi_{\rm n}(x,y,z) = - \frac{G M_{\rm n}}{\sqrt{x^2 + y^2 + z^ 2 + c_{\rm n}^2}},
  \label{Vn}
  \end{equation}
  where $G$ is the gravitational constant, while $M_{\rm n}$ and $c_{\rm n}$ are the mass and the scale length of the nucleus, respectively. At this point we would like to clarify that potential (\ref{Vn}) represents a dense and massive bulge rather than a compact object (e.g., a black hole). Therefore, relativistic effects are not taken into consideration.
  \item A rotating bar modelled by the new potential
  \begin{align}
  \Phi_{\rm b}(x,y,z) &= \frac{G M_{\rm b}}{2a}\left[\sinh^{-1} \left( \frac{x-a}{d} \right) - \sinh^{-1} \left( \frac{x+a}{d} \right) \right] = \nonumber \\
  &= \frac{G M_{\rm b}}{2a} \ln \left( \frac{x-a+\sqrt{(x-a)^2 + d^2}} {x+a+\sqrt{(x+a)^2 + d^2}} \right),
  \label{Vb}
  \end{align}
  where $d^2 = y^2 + z^2 + c_{\rm b}^2$, $M_{\rm b}$ is the mass of the bar, $a$ is the length of the semi-major axis of the bar, while $c_{\rm b}$ is its scale length (more details regarding the derivation of the new bar potential can be found in \citet{JZ15}).
  \item A flat disc described by a Miyamoto-Nagai potential \citep{MN75}
  \begin{equation}
  \Phi_{\rm d}(x,y,z) = - \frac{G M_{\rm d}}{\sqrt{x^2 + y^2 + \left(k + \sqrt{h^2 + z^ 2}\right)^2}},
  \label{Vd}
  \end{equation}
  where $M_{\rm d}$ is the mass of the disc, while $k$ and $h$ are the horizontal and vertical scale lengths of the disc, respectively.
  \item A spherically symmetric dark matter halo component using a Plummer potential
  \begin{equation}
  \Phi_{\rm h}(x,y,z) = - \frac{G M_{\rm h}}{\sqrt{x^2 + y^2 + z^ 2 + c_{\rm h}^2}},
  \label{Vh}
  \end{equation}
  where $M_{\rm h}$ and $c_{\rm h}$ are the mass and the scale length of the dark matter halo, respectively. Here it should be emphasized that a Navarro-Frenk-White (NFW) \citep{NFW96} profile or even a Hernquist profile \citep{H90} would be better for the description of the dark matter halo. However, for simplicity in the numerical calculation, we choose to use the Plummer potential.
\end{itemize}

We assume that the bar of the galaxy rotates clockwise around the vertical $z$-axis at a constant angular velocity $\Omega_{\rm b}$. On this basis, the dynamics of the galaxy are described in the corresponding rotating frame of reference where the semi-major axis of the galactic bar points into the $x$ direction, while its intermediate axis points into the $y$ direction. The total effective potential in the rotating frame of reference is
\begin{equation}
\Phi_{\rm eff}(x,y,z) = \Phi_{\rm t}(x,y,z) - \frac{1}{2}\Omega_{\rm b}^2 \left(x^2 + y^2 \right).
\label{Veff}
\end{equation}

As in Paper I, we use a system of galactic units, where the unit of length is 1 kpc, the unit of mass is $2.325 \times 10^7 {\rm M}_\odot$ and the unit of time is $0.9778 \times 10^8$ yr (about 100 Myr). The velocity unit is 10 km s$^{-1}$, the unit of angular momentum (per unit mass) is 10 km kpc s$^{-1}$, while $G$ is equal to unity. The energy unit (per unit mass) is 100 km$^2$s$^{-2}$, while the angle unit is 1 radian. In these units, the values of the involved parameters are: $M_{\rm n} = 400$ (corresponding to 9.3 $\times 10^{9}$ ${\rm M}_\odot$), $c_{\rm n} = 0.25$ kpc, $M_{\rm b} = 3500$ (corresponding to 8.13 $\times 10^{10}$ ${\rm M}_\odot$), $a = 10$ kpc, $c_{\rm b} = 1$ kpc, $M_{\rm d} = 7000$ (corresponding to 1.6275 $\times 10^{11}$ ${\rm M}_\odot$), $k = 3$ kpc, $h = 0.175$ kpc, $M_{\rm h} = 20000$ (corresponding to 4.65 $\times 10^{11}$ ${\rm M}_\odot$), $c_{\rm h} = 20$ kpc, and $\Omega_{\rm b} = 4.5$. This set of the values of the dynamical parameters defines the Standard Model (SM). The value $a = 10$ regarding the semi-major axis of the bar was also used in Paper I. Moreover in Section \ref{rs} we shall try to replicate the the spiral structure of the SBb galaxy NGC 1300, which is one of the best known barred galaxies. According to \citet{BT08} (plate 10) the semi-major axis of the bar of NGC 1300 is about 10 kpc and this fact justifies our choice. In our numerical investigation only the value of the semi-major axis of the bar will be varying in the interval $a \in [0,10]$, while the values of all the other dynamical quantities will remain constant according to SM.

The equations of motion are
\begin{align}
\dot{x} &= p_x + \Omega_{\rm b} y, \nonumber \\
\dot{y} &= p_y - \Omega_{\rm b} x, \nonumber \\
\dot{z} &= p_z, \nonumber \\
\dot{p_x} &= - \frac{\partial \Phi_{\rm t}}{\partial x} + \Omega_{\rm b} p_y, \nonumber \\
\dot{p_y} &= - \frac{\partial \Phi_{\rm t}}{\partial y} - \Omega_{\rm b} p_x, \nonumber \\
\dot{p_z} &= - \frac{\partial \Phi_{\rm t}}{\partial z},
\label{eqmot}
\end{align}
where the dot indicates the derivative with respect to the time.

In the same vein, the variational equations which govern the evolution of a deviation vector ${\vec{w}} = (\delta x, \delta y, \delta z, \delta p_x, \delta p_y, \delta p_z)$ are
\begin{align}
\dot{(\delta x)} &= \delta p_x + \Omega_{\rm b} \delta y, \nonumber \\
\dot{(\delta y)} &= \delta p_y - \Omega_{\rm b} \delta x, \nonumber \\
\dot{(\delta z)} &= \delta p_z, \nonumber \\
(\dot{\delta p_x}) &=
- \frac{\partial^2 \Phi_{\rm t}}{\partial x^2} \ \delta x
- \frac{\partial^2 \Phi_{\rm t}}{\partial x \partial y} \delta y
- \frac{\partial^2 \Phi_{\rm t}}{\partial x \partial z} \delta z + \Omega_{\rm b} \delta p_y, \nonumber \\
(\dot{\delta p_y}) &=
- \frac{\partial^2 \Phi_{\rm t}}{\partial y \partial x} \delta x
- \frac{\partial^2 \Phi_{\rm t}}{\partial y^2} \delta y
- \frac{\partial^2 \Phi_{\rm t}}{\partial y \partial z} \delta z - \Omega_{\rm b} \delta p_x, \nonumber \\
(\dot{\delta p_z}) &=
- \frac{\partial^2 \Phi_{\rm t}}{\partial z \partial x} \delta x
- \frac{\partial^2 \Phi_{\rm t}}{\partial z \partial y} \delta y
- \frac{\partial^2 \Phi_{\rm t}}{\partial z^2} \delta z.
\label{vareq}
\end{align}

The corresponding Hamiltonian (also known as the Jacobi integral of motion) which governs the motion of a test particle (star) with a unit mass $(m = 1)$ in the rotating barred galaxy model is
\begin{equation}
H = \frac{1}{2} \left(p_x^2 + p_y^2 + p_z^2 \right) + \Phi_{\rm t}(x,y,z) - \Omega_{\rm b} L_z = E,
\label{ham}
\end{equation}
where $p_x$, $p_y$ and $p_z$ are the canonical momenta per unit mass, conjugate to $x$, $y$ and $z$ respectively, $E$ is the numerical value of the Jacobi integral of motion, which is conserved, while $L_z = x p_y - y p_x$ is the angular momentum along the $z$ direction.

\begin{figure}
\begin{center}
\includegraphics[width=\hsize]{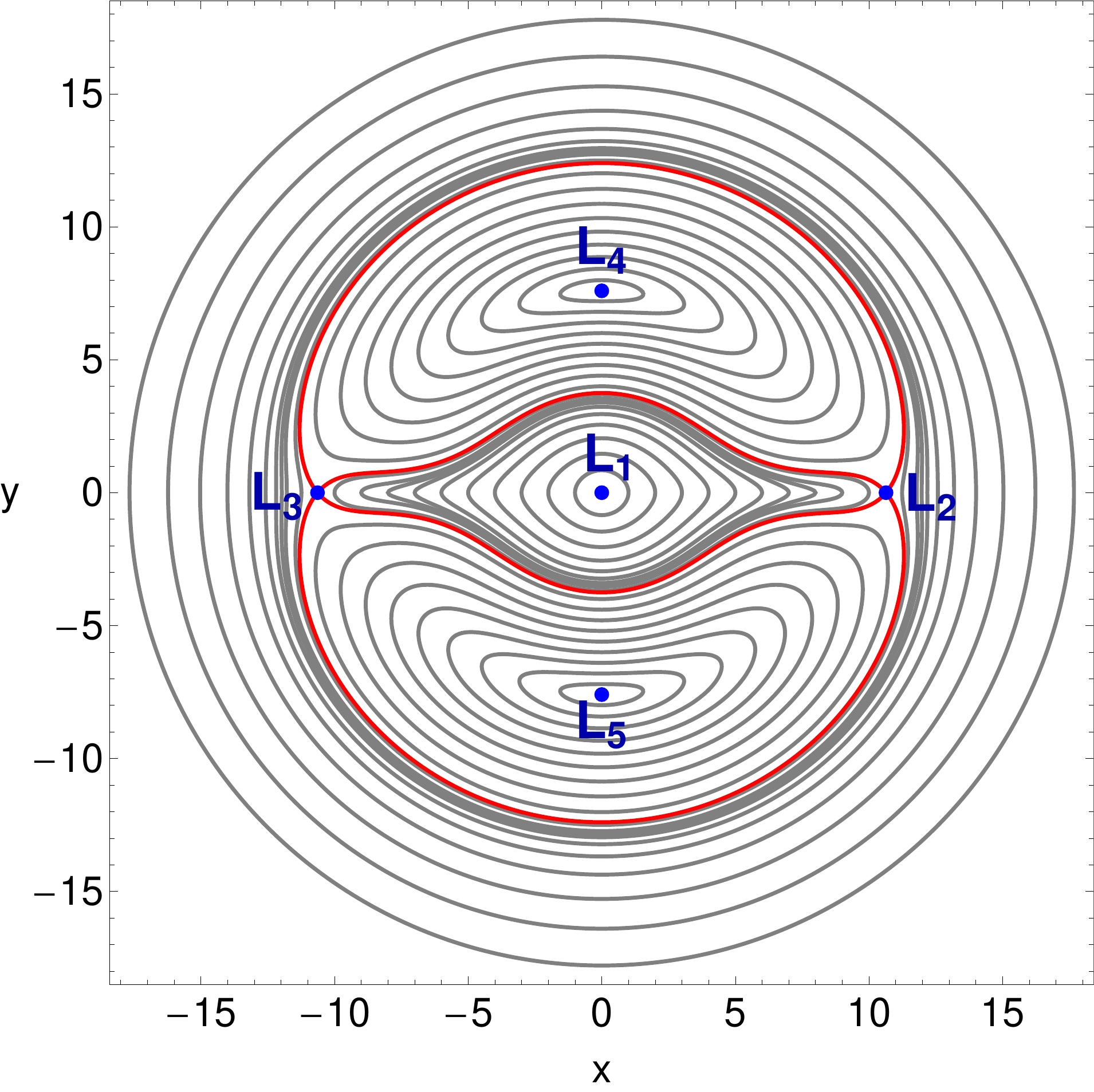}
\end{center}
\caption{The isoline contours of the effective potential on the $(x,y)$ plane (when $z = 0$) for the standard model $(a = 10)$. The positions of the five Lagrange points are indicated by blue dots. The isoline contours corresponding to the critical energy of escape $E(L_2)$ are shown in red. (For the interpretation of references to colour in this figure caption and the corresponding text, the reader is referred to the electronic version of the article.)}
\label{isoc}
\end{figure}

The Hamiltonian system of the barred galaxy has five equilibrium points (also known as Lagrange points). The coordinates of these points are the solutions of the system of differential equations
\begin{equation}
\frac{\partial \Phi_{\rm eff}}{\partial x} = \frac{\partial \Phi_{\rm eff}}{\partial y} = \frac{\partial \Phi_{\rm eff}}{\partial z} = 0.
\label{lgs}
\end{equation}
Three of the equilibrium points are located on the $x$ axis (also known as collinear points), while for the other two the $y$ coordinate has a non zero value. The central stationary point $L_1$, located at $(x,y,z) = (0,0,0)$, is a local minimum of the effective potential. The equilibrium points $L_2$ and $L_3$ are index 1 saddle points of the effective potential located at $(x,y,z) = (\pm r_L,0,0)$, where $r_L$ is the Lagrange radius. In these points the potential decreases in $x$ direction and increases in $y$ and in $z$ direction. The stationary points $L_4$ and $L_5$ on the other hand are index 2 saddle points of $\Phi_{\rm eff}$ (see Fig. \ref{isoc} for a section in the plane $z = 0$). Here the potential increases in $z$ direction and decreases in $x$ and in $y$ direction. Two important regions in position space in barred galaxies are the so-called ``region of corotation", which is defined by the circles through $L_2$, $L_3$ and $L_4$, $L_5$ and the interior region where $-r_L \leq x \leq +r_L$ \citep[see for more details][]{BT08}. In Fig. \ref{isoc} we illustrate the isoline contours of constant effective potential on the $(x,y)$ plane (when $z = 0$). The positions of the five Lagrange points are also indicated in the same figure. The numerical values of $\Phi_{\rm eff}$ at the saddle points $L_2$, $L_3$ as well as at the saddle points $L_4$ and $L_5$ are critical values of the Jacobi integral of motion. For the standard model (when $a = 10$) we have that $E(L_2) = -3242.77217493$ ($E(L_2)$ is the energy of escape) and $E(L_4) = -2800.50348529$ (remember that $E(L_2) = E(L_3)$ and $E(L_4) = E(L_5)$). When $E > E(L_2)$ the zero velocity surfaces open and two symmetrical escape channels (exits) emerge in the vicinity of the the Lagrange points $L_2$ and $L_3$. Through these channels the stars are allowed to enter the exterior region of the galaxy (when $x < -r_L$ or when $x > +r_L$) and therefore are free to escape to infinity.

The equations of motion (\ref{eqmot}) as well as the variational equations (\ref{vareq}) were integrated forward and backward in time using a double precision Bulirsch-Stoer algorithm written in standard \verb!FORTRAN 77! \citep[see e.g.,][]{PTVF92}. The adopted time step of the numerical integration was of order of $10^{-2}$ which is sufficient for the desired accuracy of our calculations. Throughout our computations the numerical error in the conservation of the Jacobi integral of motion of Eq. (\ref{ham}) was smaller than $10^{-12}$, although there were cases that the corresponding error was smaller than $10^{-14}$. All graphical illustration presented in this paper has been created using version 10.3 of Mathematica$^{\circledR}$ \citep[e.g.,][]{Wolf03}.

\section{Orbital dynamics}
\label{orbdyn}

In Paper I we investigated how the value of the semi-major axis of the bar, $a$, influences the regular or chaotic dynamics of the barred galaxy by monitoring the evolution of the orbital structure of the $(\phi, L)$ plane. In 2-dof systems the Poincar\'e map provides a very satisfying overview which regions of phase space are occupied mainly by regular motion and which regions by chaotic motion. In addition it visualizes immediately the nature of the most important structures in phase space. This is possible because for 2-dof systems the Poincar\'e map acts on a 2-dimensional domain. It would be also very interesting to reveal how the orbital structure of the 3-dof system is affected by the semi-major axis of the bar. In 3-dof systems, however, the corresponding full Poincar\'e map acts on a 4-dimensional domain and therefore it cannot be easily visualized in order to interpret the nature of the 3-dimensional orbits.

There have been several attempts to use projections, colour, multiple sections etc. to produce 2-dimensional plots of the 4-dimensional maps. Unfortunately, all these ideas fail to provide convincing results. Therefore we have to think of other possibilities, and we are aware of 2 interesting ones. First, we can look for invariant subsets of lower dimension in the phase space and restrict the Poincar\'e map to these lower dimensional sets. This idea will be followed in subsection \ref{pmap}. Second, there is another interesting alternative, the so called SALI colour-coded grids.

\begin{figure*}
\centering
\resizebox{0.95\hsize}{!}{\includegraphics{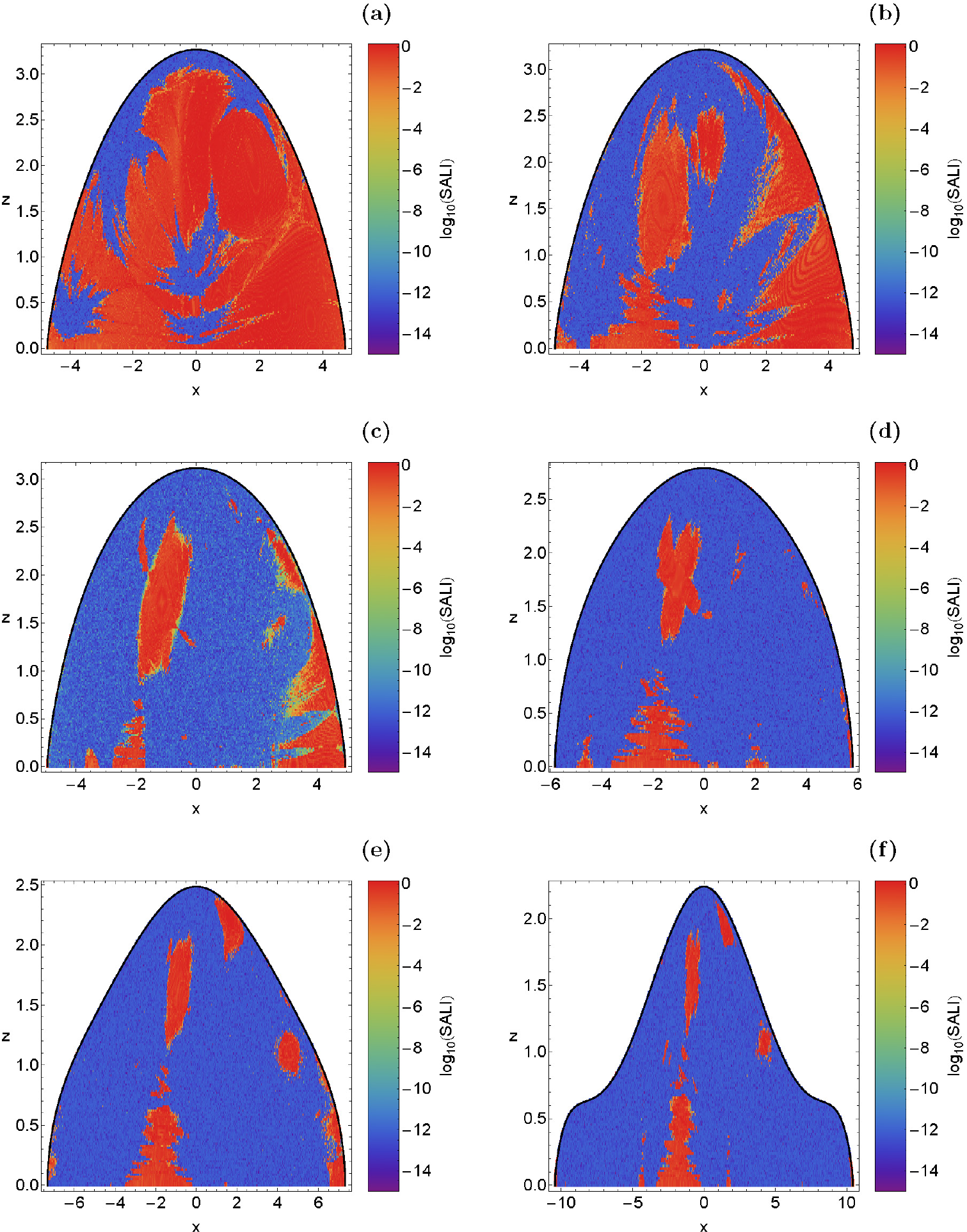}}
\caption{Regions of different values of the SALI in the corresponding dense grids of initial conditions on the $(x,z)$ plane when $E = -3245$. Light reddish colors correspond to regular motion, dark blue/purple colors indicate chaotic motion, while all intermediate colors suggest sticky orbits. The outermost black solid line is the limiting curve defined in Eq. (\ref{zvc}), while the energetically forbidden regions of motion are shown in white. (a): $a = 0.5$; (b): $a = 1.5$; (c): $a = 2.5$; (d): $a = 5.0$ (e): $a = 7.5$; (f): $a = 10$. (For the interpretation of references to colour in this figure caption and the corresponding text, the reader is referred to the electronic version of the article.)}
\label{xz}
\end{figure*}

\begin{figure}
\begin{center}
\includegraphics[width=\hsize]{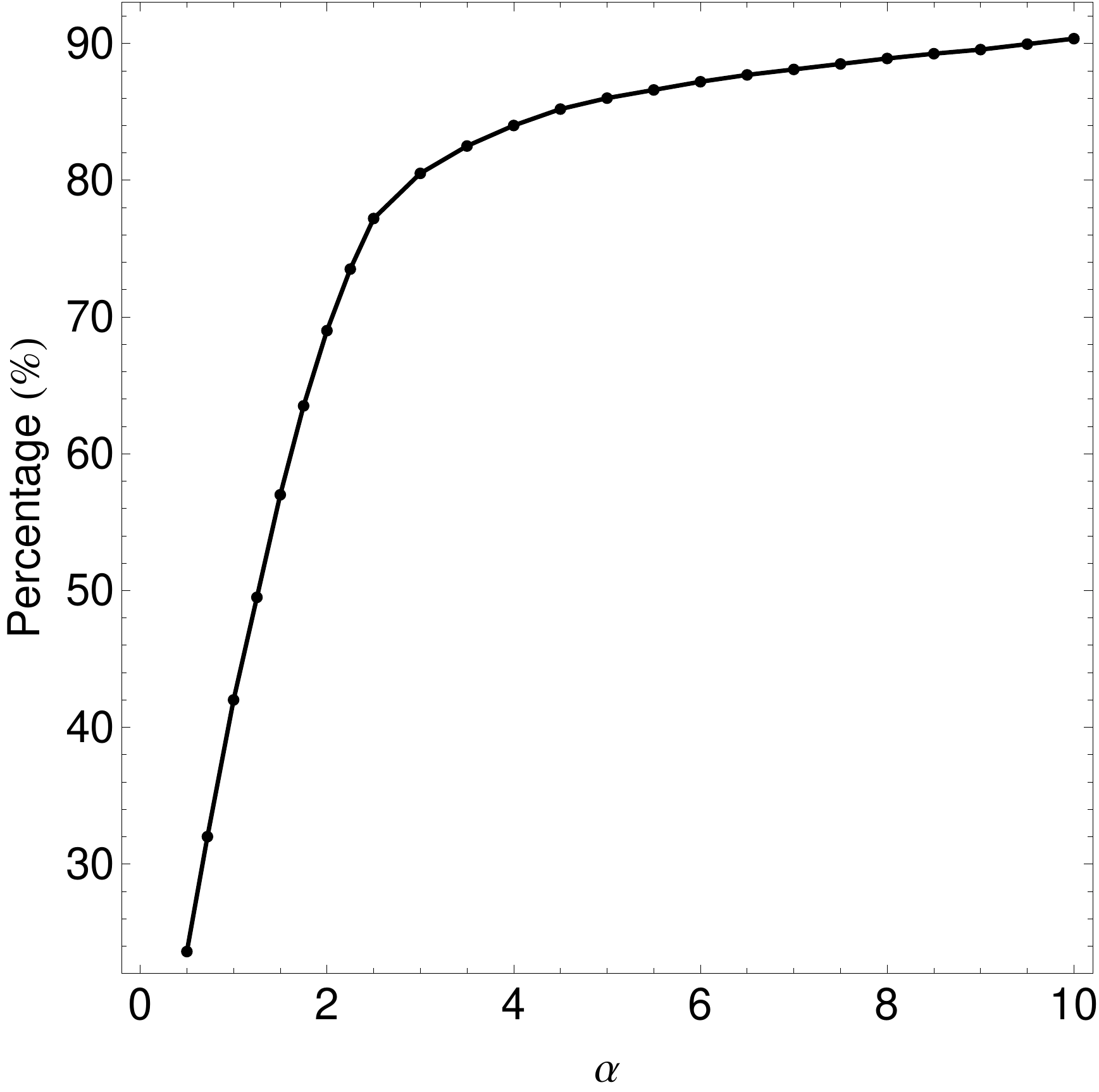}
\end{center}
\caption{Evolution of the chaotic percentage on the $(x,z)$ planes as a function of the semi-major axis $a$ of the galactic bar.}
\label{percs}
\end{figure}

Over the years, several dynamical indicators have been developed for distinguishing between order and chaos. As in Paper I, we choose to use the SALI method \citep{S01}, which has been proved a very fast and accurate tool. The mathematical definition of SALI is the following
\begin{equation}
\rm SALI(t) \equiv min(d_-, d_+),
\label{sali}
\end{equation}
where $d_- \equiv \| {\vec{w_1}}(t) - {\vec{w_2}}(t) \|$ and $d_+ \equiv \| {\vec{w_1}}(t) + {\vec{w_2}}(t) \|$ are the alignments indices, while ${\vec{w_1}}(t)$ and ${\vec{w_2}}(t)$, are two deviation vectors which initially point in two random directions. For distinguishing between ordered and chaotic motion, all we have to do is to compute the SALI along a time interval $t_{\rm max}$ of numerical integration. In particular, we track simultaneously the time-evolution of the main orbit itself as well as the two deviation vectors ${\vec{w_1}}(t)$ and ${\vec{w_2}}(t)$ in order to compute the SALI.

The time-evolution of SALI strongly depends on the particular nature of the computed orbit. More precisely, if an orbit is regular the SALI exhibits small fluctuations around non zero values, while on the other hand, in the case of chaotic orbits the SALI, after a small transient period, it tends exponentially to zero approaching the limit of the accuracy of the computer $(10^{-16})$. Therefore, the particular time-evolution of the SALI allow us to distinguish fast and safely between regular and chaotic motion. Nevertheless, we have to define a specific numerical threshold value for determining the transition from order to chaos. After conducting extensive numerical experiments, integrating many sets of orbits \citep[e.g.,][]{SABV04} we conclude that if SALI $> 10^{-4}$ the orbit is ordered, while if SALI $< 10^{-8}$ the orbit is surely chaotic. On the other hand, when the final value of SALI lies in the interval $10^{-4} \leq$ SALI $\leq 10^{-8}$ we have the case of a sticky orbit\footnote{A sticky orbit is a special type of orbit which behave as a regular one for long time intervals before it exhibits its true chaotic nature.} and further numerical integration is needed so as to fully reveal the true character of the orbit.

The basic idea for the SALI colour-coded grids is the following: We choose a 2-dimensional surface $S_I$ inside the 6-dimensional phase space and we define inside this surface a sufficiently fine grid of points. Then we can use these points as initial conditions for orbits, integrated them and therefore determine the nature of the motion for each one of these initial conditions. This approach has been successfully used in several previous works \citep[e.g.,][]{MA11,ZC13,Z14,JZ15}. Since the domains of the initial conditions are planes of dimension 2 we can present plots displaying the distribution of regular and chaotic motion. We choose the $(x,z)$ plane as the domain and we define a dense uniform grid of $1024 \times 1024$ initial conditions $(x_0,z_0)$, regularly distributed in the area allowed by the value of the total orbital energy $E$. In particular, all orbits have initial conditions $(x_0,z_0)$, $y_0 = p_{x0} = p_{z0} = 0$, while the initial value of $p_y$ is always obtained from the Jacobi integral of motion, according to Eq. (\ref{ham}) (note that we use the positive branch of the solution with $p_y > 0$). On this basis, we are able to construct again a 2-dimensional plot thus depicting the orbital structure of the $(x,z)$ plane. It should be noted that all the initial conditions of the 3-dimensional orbits lie inside the limiting curve defined by
\begin{equation}
f(x,z) = \Phi_{\rm eff}(x,y = 0,z) = E.
\label{zvc}
\end{equation}
We assign a colour to each point in $S_I$ according to the behaviour of the respective orbit, i.e. according to its numerical value of SALI at the end of the integration. The result is a 2-dimensional plot of the distribution of regular and chaotic motion in phase space. In cases where SALI plots as well as 2-dimensional Poincar\'e plots coexist, the information extracted from both methods is almost identical. A good example for the comparison of SALI plots and corresponding Poincar\'e plots is given in Fig. 4 of Paper I.

All initial conditions of the 3-dimensional orbits are numerically integrated for $10^{3}$ time units which correspond to about $10^{11}$ yr or in other words to about 10 Hubble times. This vast time of numerical integration is justified due to the presence of the sticky orbits. Therefore, if the integration interval is too short, any chaos indicator will misclassify sticky orbits as regular ones \citep[see e.g.,][]{ZC14}. In our work we decided to integrate all initial conditions of 3-dimensional orbits for a time interval of $10^{3}$ time units in order to correctly classify sticky orbits with sticky periods of at least of 10 Hubble times. At this point, it should be clarified that sticky orbits with sticky periods larger than $10^{3}$ time units will be counted as ordered ones, since such extremely high sticky periods are completely out of scope of this work.

A numerical example of SALI colour-coded grids on the $(x,z)$ plane for the 3-dof dynamics of our galaxy model is presented in Fig. \ref{xz}\footnote{We present only the $z > 0$ part of the $(x,z)$ plane because the $z < 0$ is symmetrical with respect to the $x$ axis.}. Here the energy is fixed to the value $E = -3245$, while all parameters with exception of the semi-major axis $a$ are chosen according to the standard model. The SALI value has been colour coded such that light reddish colors correspond to regular motion, dark blue/purple colors indicate chaotic motion, while all intermediate colors suggest sticky orbits. The choice of the surface $S_I$ is the following: It is a surface lying over the $(x,z)$ coordinate plane. Therefore the $x$ and $z$ coordinates are the natural coordinates in $S_I$. The sequence of plots in Fig. \ref{xz} gives a good impression how with increasing semi-major axis of the bar, i.e. with increasing strength of the bar, chaotic motion takes over in a major part of the phase space. For the numerical integration of the initial conditions of the orbits in each colour-coded grid on the $(x,z)$ plane, we needed about 1 day of CPU time on a Quad-Core i7 2.4 GHz PC. The evolution of the chaotic percentage in $S_I$ as a function of the semi-major axis of the bar is given in Fig. \ref{percs}. We observe that as the value of the semi-major axis of the bar increases the rate of chaotic orbits grows rapidly at the same time. In particular, when $a > 3$ chaotic orbits dominate the surface $S_I$ covering more than 80\% of the $(x,z)$ plane. It is interesting to note that the same behaviour (increase of the chaotic percentage as the galactic bar becomes more elongated along the $x$ direction) has been also observed in the 2-dof system investigated in Paper I (see Fig. 4).

Additional numerical calculations (not provided here) for higher values of the energy indicate that the orbital structure of the dynamical system as well as the percentages of ordered and chaotic orbits do not change significantly when the value of the energy varies in the interval $[E(L_2), -3000]$.

\section{Dynamics in the neighbourhood of the saddle points}
\label{nhims}

To understand the dynamics in the neighbourhood of the index 1 saddle point $L_2$ and in particular of the NHIM we present first an analytical perturbative treatment where we expand the effective potential around the saddle point and keep terms up to order 2 only. Later we will study numerically the effect of the higher order terms. We expand the effective potential in the form
\begin{equation}
\Phi_{\rm eff}(x,y,z) = E(L_2) - \frac{A}{2} (x - x_L)^2 + \frac{B}{2} y^2 + \frac{C}{2} z^2 + \rm h.o.t,
\end{equation}
where $x_L$ is the $x$ coordinate of the saddle point $L_2$ and for the moment we disregard the higher order terms (h.o.t). For our standard model the numerical values of the expansion coefficients are $A = 95.7815$, $B = 67.6995$, $C = 175.676$.
To insert this expansion into the Hamiltonian of Eq. (\ref{ham}), we set
\begin{equation}
\Phi_{\rm t} = \Phi_{\rm eff} + \frac{\Omega_{\rm b}^2}{2} (x^2 + y^2).
\end{equation}
Inserting the quadratic approximation for $\Phi_{\rm eff}$ gives a Hamiltonian which is a quadratic polynomial in the 6 phase space coordinates.

To simplify the treatment of the horizontal motion of the coupled $x$ and $y$ degrees of freedom we first introduce a translation of the coordinates $x$ and $p_y$ of the form
\begin{equation}
x = x_L + \tilde{x},
\end{equation}
\begin{equation}
p_y = \Omega_{\rm b} x_L + \tilde{p}_y,
\end{equation}
i.e. we move the origin of the phase space into the saddle point $L_2$. After this shift the quadratic approximation $H_2$ of the Hamiltonian has the form
\begin{align}
H_2 &= \frac{1}{2}(p_x^2 + \tilde{p}_y^2 + p_z^2) + E(L_2) - \frac{A}{2} \tilde{x}^2 + \frac{B}{2} y^2 + \frac{C}{2} z^2 \nonumber\\
&+ \frac{\Omega_{\rm b}^2}{2} ( \tilde{x}^2 + y^2 ) - \Omega_{\rm b} ( \tilde{x} \tilde{p}_y - y p_x ).
\label{ham2}
\end{align}

The corresponding linear equations of motion are
\begin{align}
\dot{\tilde{x}} &= p_x + \Omega_{\rm b} y, \nonumber \\
\dot{y} &= \tilde{p}_y - \Omega_{\rm b} \tilde{x}, \nonumber \\
\dot{z} &= p_z, \nonumber \\
\dot{p_x} &= ( A - \Omega_{\rm b}^2) \tilde{x} + \Omega_{\rm b} \tilde{p}_y, \nonumber \\
\dot{\tilde{p}}_y &= ( - B - \Omega_{\rm b}^2 ) y - \Omega_{\rm b} p_x, \nonumber \\
\dot{p_z} &= - C z.
\end{align}

If we introduce a column vector $\vec{v}$ with entries $(\tilde{x}, \tilde{p}_y, y, p_x)$ then we can write the horizontal equations in a matrix form
\begin{equation}
\frac{d}{dt} \vec{v} = M \vec{v},
\end{equation}
with the matrix $M$ given as
\begin{equation}
M = \left( \begin{array}{cccc}
0 & 0 & \Omega_{\rm b} & 1 \\
0 & 0 & -B - \Omega_{\rm b}^2 & - \Omega_{\rm b} \\
- \Omega_{\rm b} & 1 & 0 & 0 \\
A - \Omega_{\rm b}^2 & \Omega_{\rm b} & 0 & 0
\end{array} \right).
\end{equation}
Later we need the eigenvalues $\lambda_k$ of this matrix which are given by
\begin{equation}
\lambda_{1,2}^2 = D + \sqrt{D^2 + A B },
\label{eig1}
\end{equation}
\begin{equation}
\lambda_{3,4}^2 = D - \sqrt{D^2 + A B },
\label{eig3}
\end{equation}
where
\begin{equation}
D = ( A - B - 4 \Omega_{\rm b}^2 ) / 2.
\end{equation}
The right hand side of Eq. (\ref{eig1}) is always positive. Accordingly, the dynamics in the eigenplane of $M$ belonging to the real eigenvalues $\lambda_1$ and $\lambda_2$ is hyperbolic. In contrast, the right hand side of Eq. (\ref{eig3}) is
negative and therefore the eigenvalues $\lambda_3$ and $\lambda_4$ are imaginary. This implies that the dynamics in the eigenplane of $M$ belonging to $\lambda_3$ and $\lambda_4$ is elliptic.

\subsection{Fundamental periodic orbits}
\label{fpos}

We have a 3-dof system and then in the quadratic approximation of the Hamiltonian the general motion can be described as the superposition of 3 fundamental modes. In this subsection we present them. They are the vertical mode, the stable (elliptic) horizontal mode and the unstable (hyperbolic) horizontal mode.

The stable (elliptic) $z$ degree of freedom is decoupled from the other degrees of freedom and the general solution for the $z$ motion is
\begin{equation}
z(t) = \gamma \sin( \omega_z t + \phi_z),
\label{zosc}
\end{equation}
with
\begin{equation}
\omega_z = \sqrt{C}.
\end{equation}
The corresponding momentum is given as
\begin{equation}
p_z(t) = \gamma \omega_z \cos(\omega_z t + \phi_z).
\end{equation}
The energy $E_z$ in the vertical motion is given as
\begin{equation}
E_z = \frac{1}{2}\gamma^2 \omega_z^2.
\label{enez}
\end{equation}
For the stable (elliptic) horizontal mode we make the ansatz\footnote{The word ansatz is originally a German word which is now used internationally in the scientific literature and it means a functional form with which we try to construct a solution by adjusting free parameters.}
\begin{equation}
\tilde{x}(t) = \alpha \cos ( \omega_h t + \phi_h),
\label{ansx}
\end{equation}
\begin{equation}
y(t) = \beta \sin ( \omega_h t + \phi_h),
\label{ansy}
\end{equation}
where
\begin{equation}
\omega_h = \sqrt{ - \lambda_3^2} = \sqrt{ \sqrt {D^2 + AB} - D}.
\label{omh}
\end{equation}
The equations of motion give the corresponding momenta in the following form
\begin{align}
p_x(t) &= - ( \alpha \omega_h + \beta \Omega_{\rm b} ) \sin ( \omega_h t + \phi_h) \nonumber\\
&= - ( \omega_h \alpha / \beta + \Omega_{\rm b} ) y(t),
\label{anpx}
\end{align}
\begin{align}
\tilde{p}_y(t) &= ( \beta \omega_h + \alpha \Omega_{\rm b} ) \cos ( \omega_h t + \phi_h) \nonumber\\
&= ( \omega_h \beta / \alpha + \Omega_{\rm b} ) \tilde{x} (t).
\label{anpy}
\end{align}
We introduce $\delta = \alpha / \beta$. The ansatz is a solution of the equations of motion if $\delta$ fulfils
\begin{equation}
\delta = - 2 \Omega_{\rm b} \omega_h/(\omega_h^2 + A).
\label{ansd}
\end{equation}
An equivalent condition is
\begin{equation}
\delta^2 = (\omega_h^2 - B)/(\omega_h^2 +A).
\label{ansd2}
\end{equation}
To get the amplitudes $\alpha$ and $\beta$ themselves, we need the energy $E_h$ in the horizontal motion around $L_2$. For the case which we will be mainly interested in, namely for a combination of the vertical mode with the stable horizontal mode, but no motion in the unstable horizontal mode, it is given as
\begin{equation}
E_h = E - E_z - E(L_2).
\label{eneh0}
\end{equation}
Then $\beta$ is given by
\begin{equation}
\beta^2 = 2 E_h / ( \omega_h^2 - \delta^2 A),
\end{equation}
and $\alpha$ is given by
\begin{equation}
\alpha = \delta \beta.
\end{equation}
The energy of the horizontal motion expressed by the amplitude of the horizontal motion can also be given in the following forms
\begin{align}
E_h &= \frac{\alpha^2}{2} \omega^2_ h + \frac{\beta^2}{2} B =
\frac{\beta^2}{2} \omega^2_h - \frac{\alpha^2}{2} A \nonumber\\
&= \frac{\alpha^2}{2} \; \; \frac{\omega^4_h + AB}{\omega^2_h -B} =
\frac{\beta^2}{2} \; \; \frac{\omega^4_h + AB}{\omega^2_h +A}.
\label{eneh}
\end{align}

All orbits treated so far are unstable in the eigenplane of the matrix $M$ belonging to the eigenvalues $\lambda_1$ and $\lambda_2$ where the instability exponent is $\lambda_1$. In the limit $E \to E(L_2)$ the vertical periodic orbit constructed in this subsection represents the vertical Lyapunov orbit which we will call $\Gamma_v$ in the following, and
the horizontal periodic orbit represents the horizontal Lyapunov orbit which we will call $\Gamma_h$ in the following. When we insert the numerical values for the parameters $A$, $B$, $C$, $\Omega_{\rm b}$ into the equations of this subsection then we obtain perfect coincidence with the numerical properties of the Lyapunov orbits in the limit $E \to E(L_2)$.

Finally let us present the unstable (hyperbolic) horizontal mode, its orbits are given as
\begin{equation}
\tilde{x}(t) = a ( \exp ( \lambda_1 t)+c\exp(-\lambda_1 t)),
\end{equation}
\begin{equation}
y(t) = b ( \exp ( \lambda_1 t)-c\exp(-\lambda_1 t)),
\end{equation}
\begin{equation}
p_x(t)=(\lambda_1 a/b - \Omega_{\rm b}) y(t),
\end{equation}
\begin{equation}
\tilde{p}_y (t) = (\lambda_1 b/a + \Omega_{\rm b}) \tilde{x}(t),
\end{equation}
where
\begin{equation}
d = a/b = 2 \Omega_{\rm b} \lambda_1 / ( \lambda_1^2 - A).
\end{equation}

\subsection{The NHIM over the saddle $L_2$ in quadratic approximation}
\label{quad}

Let us now consider all possible orbits with no motion at all in the unstable mode and for the moment still in the quadratic approximation of the effective potential. Because of the decoupling of the vertical $z$ motion from the
horizontal motion on the $(x,y)$ plane all such orbits are the quasi-periodic superposition of the periodic $z$ motion and the periodic stable horizontal motion studied in the previous subsection. Let us assume that the total energy is $E$. Then the energy available for the motion in the neighbourhood of the saddle $L_2$ is $E_r = E - E(L_2)$. This available energy can be split between the decoupled vertical motion and the stable horizontal motion such that $E_h + E_z = E_r$. This gives a 1-dimensional continuum of possibilities. In addition we have the freedom to choose for the relative phase shift $\phi_h - \phi_z$ between horizontal and vertical motion any value between 0 and $2 \pi$. This is another 1-dimensional continuum of possibilities. Accordingly in total we have a 2-dimensional continuum of orbits which stay over the saddle region permanently and do not move away along the unstable directions neither in the future nor in the past. The set of all these particular orbits forms a 3-dimensional surface which we call $S_E$. As we will see in a moment, it has the topology of the 3-dimensional sphere $S^3$ in the 5-dimensional energy shell. It implies a corresponding 2-dimensional invariant surface in the 4-dimensional domain of the Poincar\'e map for a fixed energy. All orbits belonging to $S_E$ are neutrally stable (parabolic) in tangential direction to $S_E$ and hyperbolic in normal direction. All these properties together show that the surface $S_E$ is a NHIM of codimension 2. The motion normal to the NHIM is hyperbolic and therefore the NHIM has stable and unstable manifolds of codimension 1. The property of these stable and unstable manifolds to be of codimension 1 is important. It shows that these surfaces are dividing the phase space and are able to direct and channel the general flow over the saddle points.

The topology of the NHIM for fixed energy $E > E(L_2)$ can be understood as follows. Eqs. (\ref{ansx}) and (\ref{ansy}) show that the stable horizontal motion consists of a set of concentric ellipses in the position space where the axis ratio is
the same for all these ellipses and is given by Eq. (\ref{ansd}). The size of the ellipses is related to the horizontal energy by Eq. (\ref{eneh}). For a given value of the total energy $E$ there is a maximal value of $E_h$ and therefore also a
maximal value of the size of the elliptic orbit in the horizontal position space. Accordingly the horizontal orbits fill an elliptical disc $DC_E$ in position space. Any one of these horizontal orbits in position space is lifted into the phase space by Eqs. (\ref{anpx}) and (\ref{anpy}) which establish linear relations between the horizontal coordinates and the horizontal momenta. The lift of the disc $DC_E$ into the phase space is given by the same equations and results in an elliptical disc $DP_E$. To get the full motion we place now over each point of $DP_E$ a 1-dimensional fiber which consists of an orbit of the $z$ motion where the amplitude of the $z$ motion is chosen according to the Eqs. (\ref{enez}), (\ref{eneh0}) and (\ref{eneh}). For all points in the interior of $DP_E$ the attached fiber has the topology of a circle, only for points on the boundary of $DP_E$ this circle shrinks to a single point. This is the description of $S_E$ in terms of orbits.

The two relations of Eqs. (\ref{anpx}) and (\ref{anpy}) between horizontal position and momentum define a 4-dimensional plane $SP$ in the 6-dimensional phase space. To write the conditions for points of $S_E$ in the form of equations, we use Eqs. (\ref{anpx}) and (\ref{anpy}) to eliminate the coordinates $p_x$ and $\tilde{p}_y$ in Eq. (\ref{ham2}). Thereby we construct a function $H_{2,r}(\tilde{x},y,z,p_z)$, it is the quadratic approximation of the energy function restricted to the plane $SP$ and it is given as
\begin{align}
H_{2,r}(\tilde{x},y,z,p_z) &= \frac{1}{2}p_z^2 + \frac{C}{2} z^2 + E(L_2) + \frac{1}{2}\tilde{x}^2 ( \omega_h^2 \beta^2 / \alpha^2 -A) \nonumber\\
&+ \frac{1}{2} y^2 ( \omega_h^2 \alpha^2 / \beta^2 + B ).
\label{h2r}
\end{align}
Finally we can describe $S_E$ in the following form as the intersection between the energy shell and the plane $SP$
\begin{align}
S_E &= \{ (\tilde{x},y,z,p_x,\tilde{p}_y,p_z) \mid  p_x = - ( \omega_h \alpha / \beta + \Omega_{\rm b} ) y , \tilde{p}_y \nonumber\\
&= (\omega_h \beta / \alpha + \Omega_{\rm b} ) \tilde{x}, \ H_{2,r}(\tilde{x},y,z,p_z) = E \}.
\label{sep}
\end{align}
Inserting Eq. (\ref{ansd}) or (\ref{ansd2}) into Eq. (\ref{h2r}) and using Eq. (\ref{omh}) shows that the coefficient
of $\tilde{x}^2$ in Eq. (\ref{h2r}) is always positive and that therefore the condition $H_{2,r} = E$ in Eq. (\ref{sep}) defines an ellipsoid. In total we see that in the quadratic approximation the NHIM surface $S_E$ is a 3-dimensional ellipsoid sitting in the 5-dimensional energy surface of the phase space. Then also for small perturbations the NHIM has the topology of a 3-dimensional sphere $S^3$. This is the standard structure of a NHIM over an index 1 saddle of a 3-dof system when the energy is above but close to the saddle energy \citep[see e.g.,][]{MS14}.

Because the NHIM is invariant we can construct the restriction of the Poincar\'{e} map to the NHIM. In the quadratic approximation it is a rather simple map. We use the intersection condition $z = 0$ and then in the harmonic approximation studied at the moment the Poincar\'e map becomes a stroboscopic map for the horizontal motion where the return time is $T_z = 2 \pi / \omega_z$. During this time the phase of the elliptic horizontal motion advances by $\Delta \phi_h = 2 \pi \omega_h / \omega_z $. In total the restricted Poincar\'e map becomes the rotation of a disc in analogy to the Poincar\'e map of a 2-dof anisotropic harmonic oscillator. The distribution of the available energy between the vertical and the horizontal degrees of freedom is constant along the invariant curves of the restricted map, i.e. the energy in the vertical motion and the energy in the horizontal motion are conserved separately. The quadratic approximation is not able to describe the bifurcations of the Lyapunov orbits under an increase of the energy.

\begin{figure*}
\centering
\resizebox{\hsize}{!}{\includegraphics{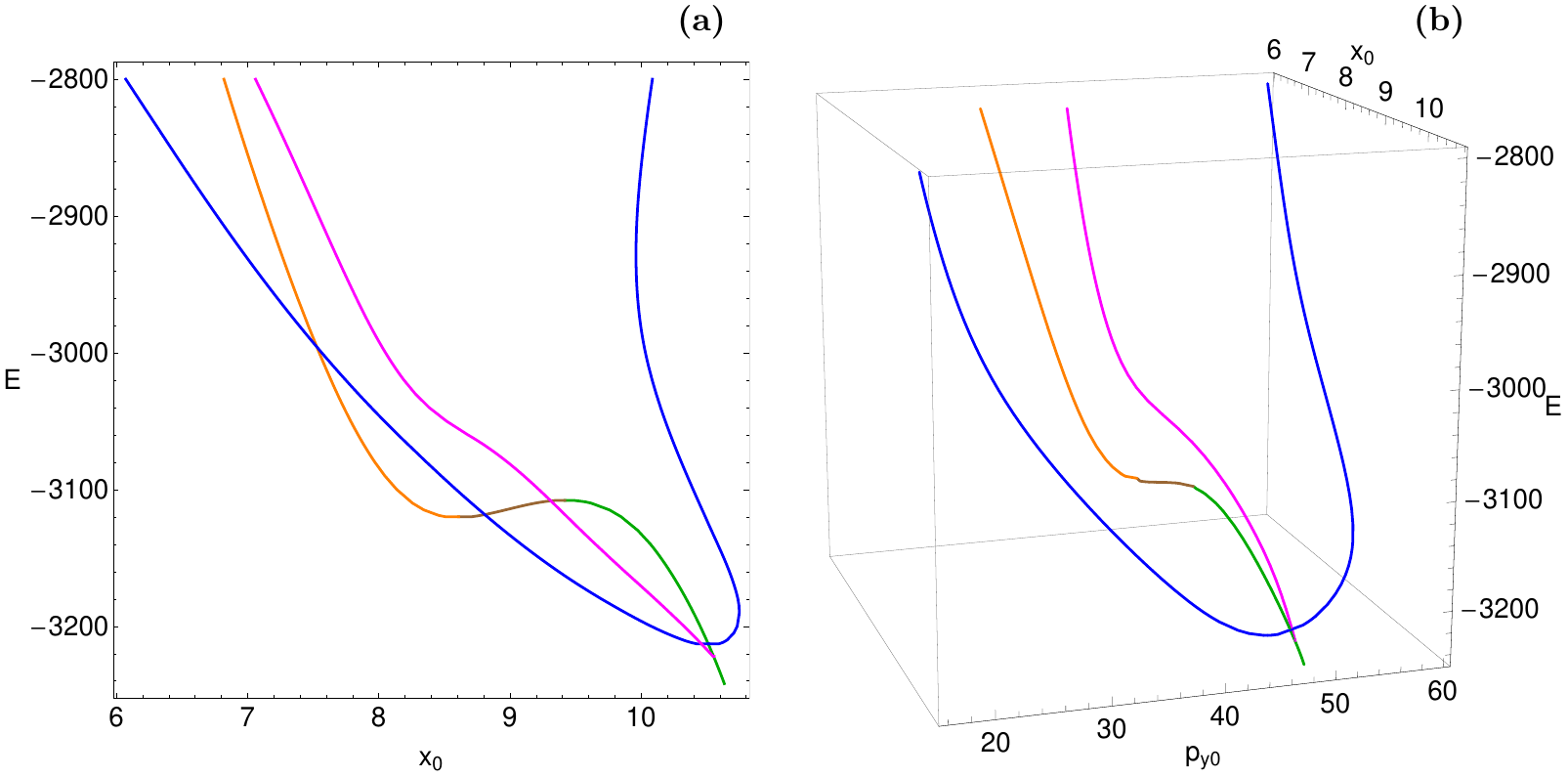}}
\caption{(a-left): Evolution of the $x_0$ initial conditions of the periodic orbits as a function of the orbital energy $E$. (b-right): Evolution of the $x_0$ and $p_{y0}$ initial conditions of the periodic orbits as a function of the orbital energy $E$). The colour code is the following: the vertical Lyapunov orbit $\Gamma_v$ (green); the first continuation of the Lyapunov orbit $\Gamma_v$, which is the orbit $\Gamma_c$ (brown); the second continuation of the orbit $\Gamma_v$, which is the orbit $\Gamma_d$ (orange); the tilted loop orbit from first pitchfork bifurcation (magenta); the tilted orbit from the second pitchfork bifurcation (blue). (For the interpretation of references to colour in this figure caption and the corresponding text, the reader is referred to the electronic version of the article.)}
\label{fpo}
\end{figure*}

Now the important question arises: What happens to the NHIM when we include the higher order terms into the effective potential and into the equations of motion? In general the restriction of the map to the NHIM will become more complicated, the dynamics will also develop tangential instability. However, when the perturbation is sufficiently small (i.e. the energy is still close to the saddle value $E(L_2)$) then the tangential instabilities are still small compared to the normal instability and then the persistence theorem of NHIMs guarantees the survival of the NHIM under the perturbation. It may be
deformed smoothly and displaced a little, but it remains an invariant surface of codimension 2 and also its stable and unstable manifolds survive. Of course, for large perturbations the tangential instability of the NHIM can become large or the normal hyperbolicity may be lost and then the NHIM may change its qualitative structure and may decay. The reader can find more information regarding the persistence theorem and the bifurcations of NHIMs in \citet{BB13,E13,F71,W88,W94} and \citet{AB12,MS14,MCE13,LST06,TTK11,TTK15,TTT15}, respectively.

\subsection{The development scenario of the Lyapunov orbits}
\label{lyap}

In this subsection we present numerical results for the development scenario of the Lyapunov orbits. The Lyapunov orbit $\Gamma_h$ follows a rather simple scenario. The normal instability decreases monotonically with increasing energy and at the rather high energy value of -2678 it becomes normally stable in a pitchfork bifurcation where it splits off two new horizontal periodic orbits. The bifurcation diagram of the Lyapunov orbit $\Gamma_h$ is given in Fig. 11 of Paper I. The trace given in part (b) of the same figure is the normal trace according to our present terminology explained in detail below. Plots of the Lyapunov orbit $\Gamma_h$ and of the split off orbits for energy -2600 have been given in Fig. 12 of Paper I. Therefore we dedicate the rest of this subsection to the more complicated scenario of the vertical Lyapunov orbit $\Gamma_v$. When we use the expression normal directions or tangential directions in the following then it always refers to the NHIM surface.

The vertical Lyapunov orbit $\Gamma_v$ is born unstable in normal direction and stable in tangential direction. Near the energy -3223 it suffers a first pitchfork bifurcation where it splits off two tangentially stable orbits and the orbit $\Gamma_v$ itself becomes tangentially unstable. At the energy -3214, the orbit $\Gamma_h$ suffers a second pitchfork bifurcation where it splits off two tangentially unstable orbits and it returns to tangential stability. Next at energy -3108 it collides in normal direction with another periodic orbit (called orbit $\Gamma_c$ in the following) which is tangentially stable and also normally stable. Orbits $\Gamma_v$ and $\Gamma_c$ disappear in a saddle centre bifurcation. The orbit $\Gamma_c$ itself is created in a saddle centre bifurcation at energy -3120 together with a further periodic orbit (called orbit $\Gamma_d$ in the following) which is unstable in normal direction. In total the effect of these two saddle centre bifurcations is to replace orbit $\Gamma_v$ by the very similar orbit $\Gamma_d$ which only lies at smaller values of the $x$ coordinate. The 4 orbits split off in the two pitchfork bifurcations are tilted loop orbits.

In the two parts of Fig. \ref{fpo} we present this bifurcation scenario of the Lyapunov orbit $\Gamma_v$ graphically. In panel (a) we show for all the seven involved orbits the evolution of the $x$ coordinate in the Poincar\'e map (i.e. the $x$ coordinates of these orbits at the moment of positive intersection with the plane $z = 0$) as function of the energy. The green branch is the orbit $\Gamma_v$, the brown branch is the orbit $\Gamma_c$, the orange branch is the orbit $\Gamma_d$, the magenta branch represents the two orbits split off in the first pitchfork bifurcation of the Lyapunov orbit $\Gamma_v$ (the $x$ coordinates of both of them are equal) and the blue curve represents the two orbits split off in the second pitchfork bifurcation of the orbit $\Gamma_v$. This diagram gives an impression in which way the orbit $\Gamma_d$ takes over the role of the orbit $\Gamma_v$.

\begin{figure*}
\centering
\resizebox{\hsize}{!}{\includegraphics{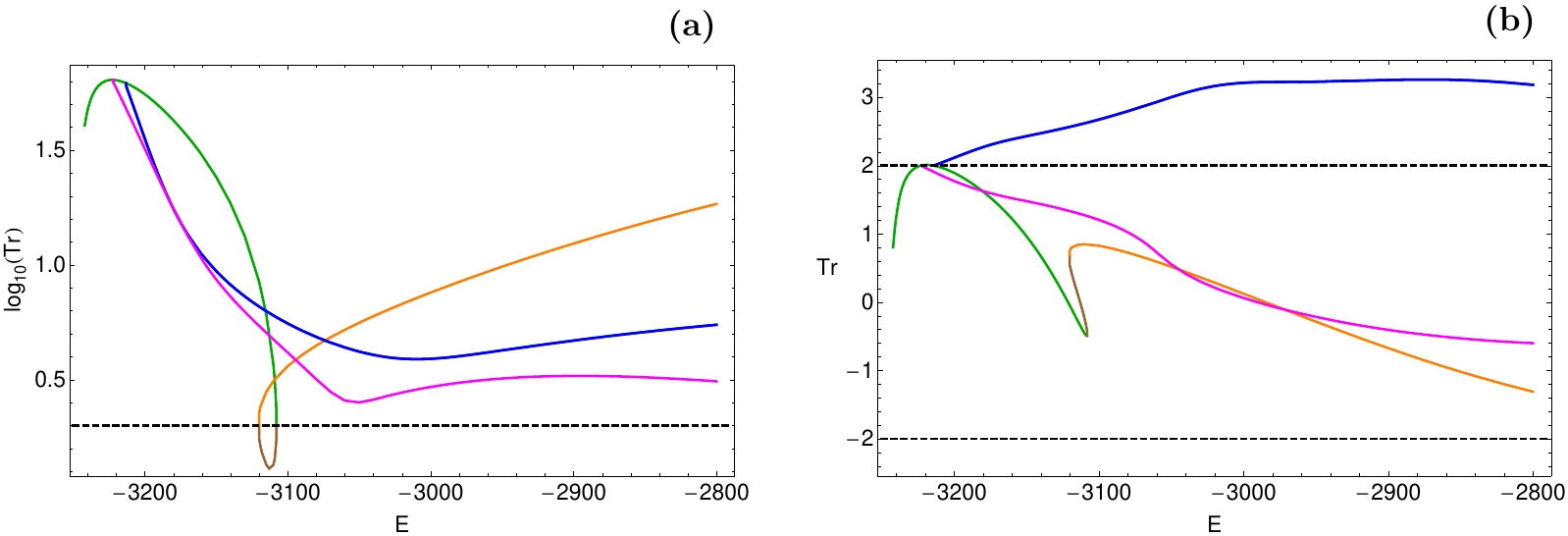}}
\caption{Evolution of (a-left): the normal and (b-right): the tangential traces of the monodromy matrix of the families of periodic orbits as a function of the total orbital energy $E$. The colour code is the same as in Fig. \ref{fpo}. The horizontal black dashed lines correspond to the critical values $\rm Tr_{\rm crit} = \pm 2$, which distinguish between stable and unstable motion. (For the interpretation of references to colour in this figure caption and the corresponding text, the reader is referred to the electronic version of the article.)}
\label{trs}
\end{figure*}

All orbits involved in this bifurcation scenario have monodromy matrices with a natural splitting into two real $2 \times 2$ blocks normal and tangential to the NHIM surface. Complex spiralling does never occur in the present scenario. Therefore we  present in Fig. \ref{trs}(a-b) the normal and the tangential stability properties of the seven periodic orbits involved in this development scenario by showing the traces of these two blocks as function of the energy. Note that the two orbits split off in a pitchfork bifurcation have exactly the same stability properties because of symmetry reasons. Therefore we see only five curves in the two parts of Fig. \ref{trs}, but they represent all seven orbits involved. Panel (a) of Fig. \ref{trs} shows the normal stability traces. We see that all orbits with exception of the orbit $\Gamma_c$ are normally unstable and therefore qualify as possible parts of the NHIM surface. Only the orbit $\Gamma_c$ as stable in normal direction can not be part of the NHIM surface and this has interesting consequences explained below. Panel (b) of Fig. \ref{trs} shows the tangential stability traces. The orbits split off in the second pitchfork bifurcation of the Lyapunov orbit $\Gamma_v$ are tangentially unstable up to the rather high value -2547 of the energy and orbit $\Gamma_v$ is tangentially unstable in the short energy interval between -3223 and -3214, i.e. between the two pitchfork bifurcations. The rest of the involved orbits is tangentially stable.

\begin{figure*}
\centering
\resizebox{\hsize}{!}{\includegraphics{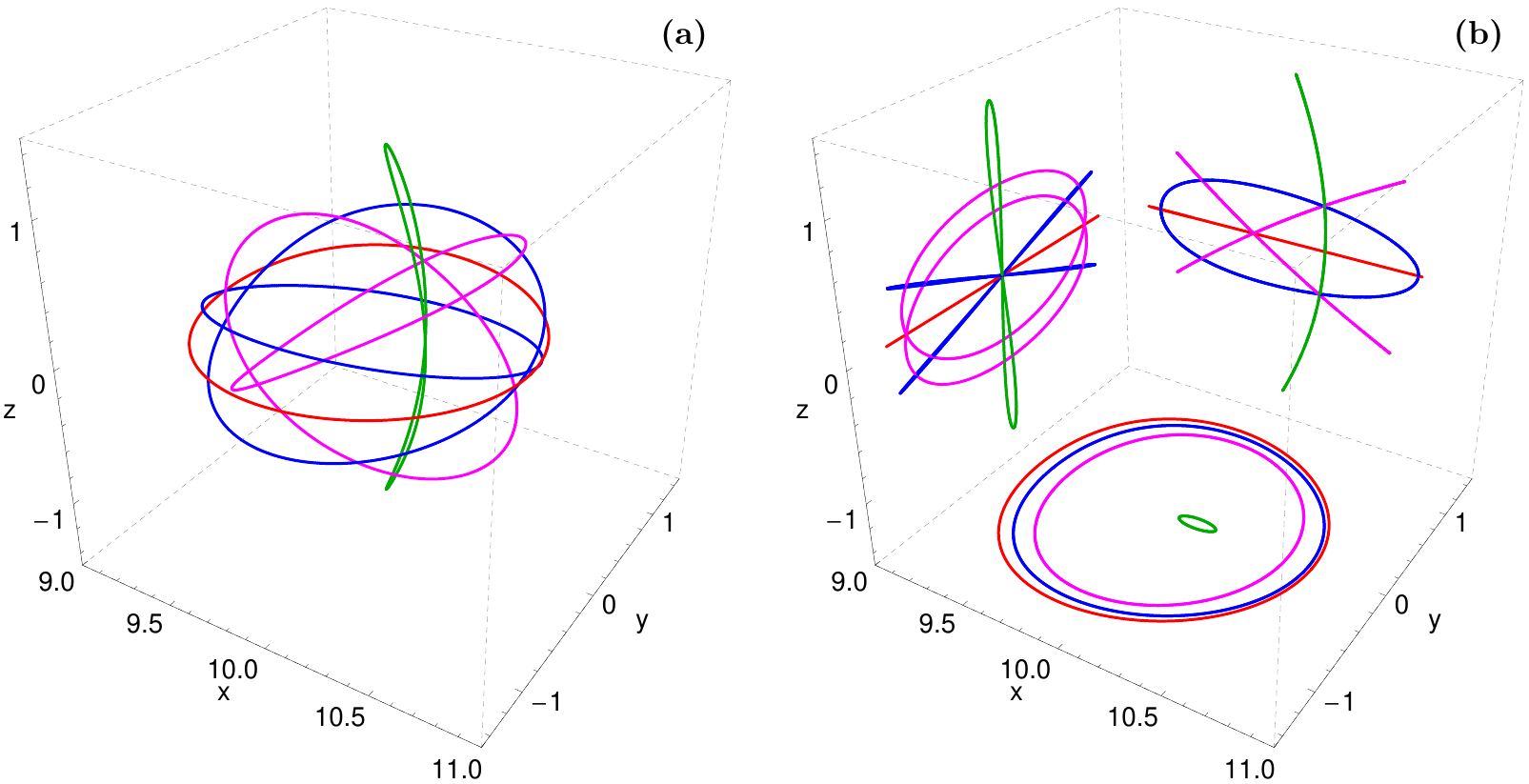}}
\caption{(a-left): The two Lyapunov orbits (horizontal and vertical) and the four tilted orbits in the configuration $(x,y,z)$ space, for the energy $E = -3150$. (b-right): The projections of the six periodic orbits into the primary planes $(x,y)$, $(x,z)$ and $(y,z)$. The colour code is the same as in Fig. \ref{fpo}, while the horizontal Lyapunov periodic orbit $\Gamma_h$ is plotted in red. (For the interpretation of references to colour in this figure caption and the corresponding text, the reader is referred to the electronic version of the article.)}
\label{orbs}
\end{figure*}

In Fig. \ref{orbs}(a-b) we present the two Lyapunov orbits (horizontal and vertical) and the four tilted loop orbits in position space for the energy $E = -3150$. Note how the reflection symmetry in $z$ of the dynamics acts on the important periodic orbits. As point sets the Lyapunov orbits $\Gamma_h$ and $\Gamma_v$ are both invariant under the reflection $z \rightarrow -z $. In contrast, the individual tilted loop orbits are not invariant under this reflection. Instead one copy is mapped into the other one. This behaviour is typical for pairs of orbits created in pitchfork bifurcations. For detailed explanations on pitchfork bifurcations see section 3.4 in \citet{GH83} or sections 20.1e and 21.1c in \citet{W03}.

\subsection{Numerically constructed restricted Poincar\'e map on the NHIM}
\label{pmap}

The best graphical presentation of the whole development scenario found in the previous subsection is given by a sequence of plots of the Poincar\'e map restricted to the NHIM. It displays at the same time the development scenario of the NHIM itself. Numerically this map is constructed as explained in \citet{GDJ14} as a combination of the 4-dimensional Poincar\'e map with a projection on the stable manifold of the NHIM. The inclusion of this projection can be interpreted as a version of the control of chaos where we keep the numerical orbit in the neighbourhood of an unstable invariant subset \citep[see e.g.,][]{OGY90,SOGY90}. The Lyapunov orbit $\Gamma_h$ lies completely in the intersection plane $z = 0$ of the map. Therefore, $\Gamma_h$ is the energetic boundary of the domain of the restricted map. If we want to represent also the orbit $\Gamma_h$ as a single fixed point in the restricted map then we can contract it, i.e. contract the boundary of the domain, to a single point. Thereby the domain of the restricted map, i.e. the NHIM surface in the Poincar\'e map, acquires the topology of a sphere $S^2$. This is very similar to the contraction of the boundary in the NHIM construction for the example of an electron in a perturbed magnetic dipole field as explained in \citet{GJ15}. To show plots of this restricted map we have to project the NHIM surface in some form. A relatively simple possibility is to project into the $(\phi, L)$ plane\footnote{Remember that $\phi = \arctan(y/x)$ and $L = x p_y - y p_x$, as usual.}. For energies close to the saddle energy this projection is 1:1. Unfortunately for higher energies this property is lost, however it is equally lost in other projections.

\begin{figure*}
\centering
\resizebox{0.9\hsize}{!}{\includegraphics{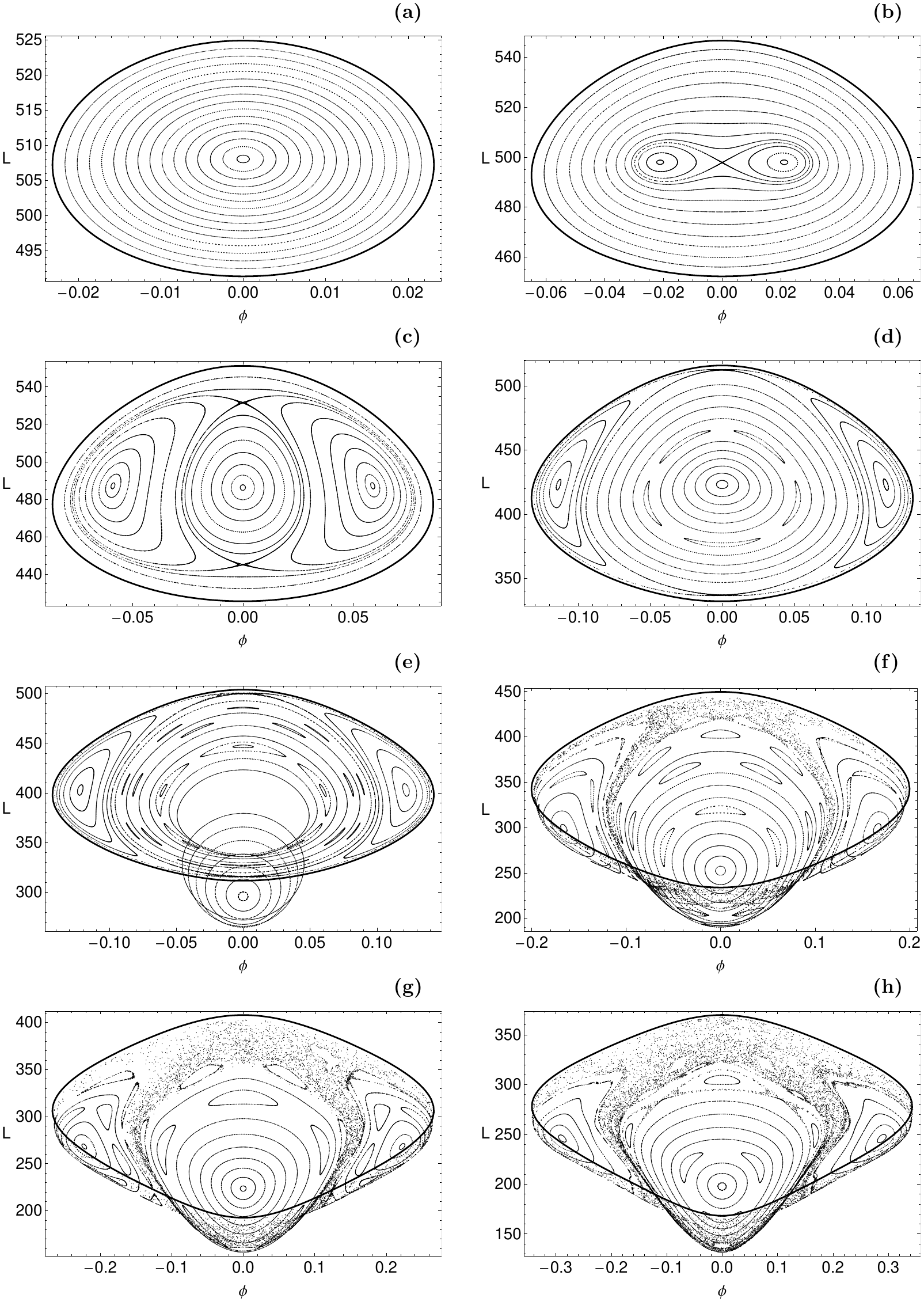}}
\caption{Projections of the NHIM surfaces into the $(\phi, L)$ plane. The outermost solid closed curve corresponds to the horizontal Lyapunov orbit. (a): $E = -3240$; (b): $E = -3220$; (c): $E = -3200$; (d): $E = -3120$; (e): $E = -3100$; (f): $E = -3000$; (g): $E = -2900$; (h): $E = -2800$.}
\label{maps}
\end{figure*}

\begin{figure*}
\centering
\resizebox{0.7\hsize}{!}{\includegraphics{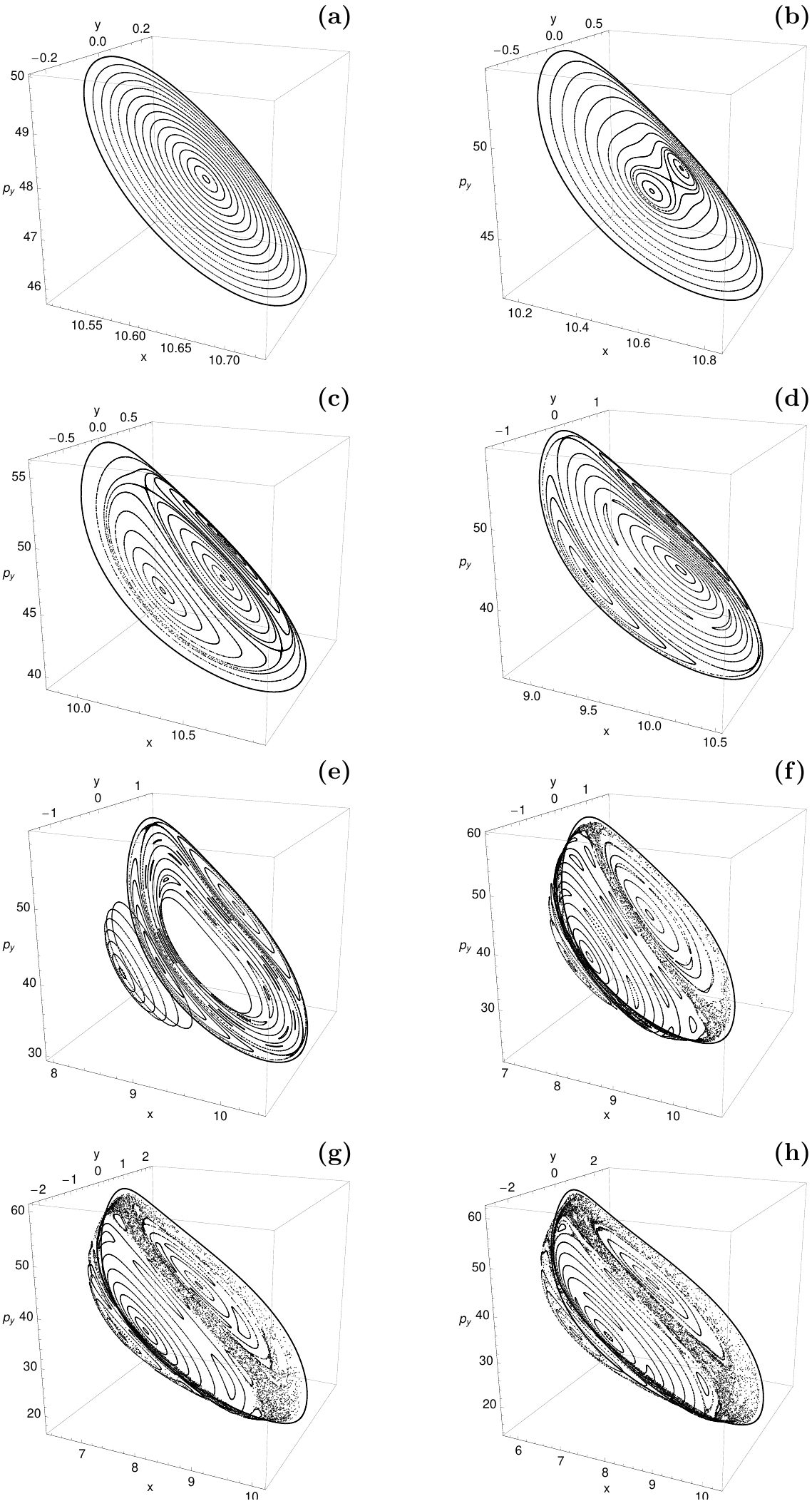}}
\caption{Projections of the NHIM surfaces into the $(x,y,p_y)$ space. The outermost solid closed curve corresponds to the horizontal Lyapunov orbit. (a): $E = -3240$; (b): $E = -3220$; (c): $E = -3200$; (d): $E = -3120$; (e): $E = -3100$; (f): $E = -3000$; (g): $E = -2900$; (h): $E = -2800$.}
\label{maps3}
\end{figure*}

The numerical maps are plotted in Fig. \ref{maps}(a-h) for eight values of the energy. Panel (a) presents the energy -3240. Here the numerical map comes very close to the one described above for the quadratic approximation. The central fixed point represents the Lyapunov orbit $\Gamma_v$. The invariant circles around it represent quasi-periodic motion which is a superposition of the vertical and the horizontal periodic motion. In the centre, all the available energy $E_r$ is in the vertical motion. Going further out more and more energy goes into the horizontal motion until at the boundary all available energy $E_r$ is in the horizontal motion. Along all the invariant curves the distribution of the available energy between the vertical motion and the horizontal motion is approximately constant for this energy close to the saddle energy.

Panel (b) of Fig. \ref{maps} gives the numerical map for energy -3220. Here the first pitchfork bifurcation of the orbit $\Gamma_v$ has already occurred. Correspondingly the central fixed point in the map has become unstable and has split off a pair of stable fixed points. The unstable fixed point has turned into the centre of a very fine chaos strip which looks in the plot like a separatrix curve. When we proceed to panel (c) of Fig. \ref{maps} for energy -3200 then the second pitchfork bifurcation of the orbit $\Gamma_v$ has occurred, the central fixed point in the map has returned to stability and has split off a pair of unstable fixed points which have taken over the role of the centres of the fine chaos strip which still looks very close to a separatrix curve. The structure created by the two consecutive pitchfork bifurcations looks exactly the same as a secondary island chain belonging to a 1:1 resonance where the number of elliptic and hyperbolic points is doubled from 1 to 2 because of a discrete symmetry.

In panel (d) of Fig. \ref{maps} we have arrived at energy -3120. In the development scenario between parts (c) and (d) no qualitative large scale changes have occurred. We only observe a smooth change of large scale structures. For energy -3120 the perturbation is already stronger and therefore the first secondary structures have become visible. The most important one is an island chain of period 5. When we now proceed to the energy -3100 shown in panel (e) of Fig. \ref{maps}, then an extremely drastic change happens. As we have already seen before at energy -3108 the orbit $\Gamma_v$ is lost. This cuts a hole into the NHIM surface. We have also seen that the orbit $\Gamma_v$ is replaced by the orbit $\Gamma_d$ located further inside. Then a new piece of invariant surface is created around the orbit $\Gamma_d$. Therefore panel (e) of Fig. \ref{maps} shows the superposition of 2 disjoint pieces of invariant surface. Of course, in the 4-dimensional domain of the full Poincar\'e map these two pieces do not overlap. Only the projection into the $(\phi, L)$ plane shows apparent intersections. Numerical behaviour indicates that the normal instability of the invariant surface diminishes when approaching the boundary of the hole in the original NHIM surface and also when approaching the boundary of the new piece of NHIM around the orbit $\Gamma_d$.

In the next panel (f) of Fig. \ref{maps} the energy is -3000. In the development from panel (e) to panel (f) something amazing happens which we do not yet fully understand. The two pieces of the NHIM surface join again to a single connected surface having the same topology as the NHIM for low energies. For this energy the separatrix has already turned into a large scale chaos strip indicating that we now enter the region of mid-size perturbation. In the next panel (g) of Fig. \ref{maps} the energy is increased to the value -2900. Compared to panel (f) we do not see any qualitative changes of large scale structures. Only the large chaotic sea grown out of the separatrix has become larger and starts to disintegrate from the outside the island structures around the elliptic fixed points. Finally in panel (h) of Fig. \ref{maps} the energy has been increased to the value -2800. Here we are very close to the critical energy $E(L_4)$ of the Lagrange points $L_4$ and $L_5$. However, the NHIM under study comes nowhere close to these points and therefore it is not affected by the corresponding change of the global topology of the accessible part of the position space. For still higher energy the complete plane $z = 0$ is accessible and there are no more any bottle necks related to the Lagrange points $L_2$ and $L_3$. Therefore the further development of the NHIM for still higher energy is no longer of relevance for the topic of the present article, namely for the escape dynamics over the index 1 saddle points of the effective potential.

Accordingly, we do not discuss in detail the development scenario of the NHIM for still higher values of the energy. We give just some short remarks: For $E = -2678$ the Lyapunov orbit $\Gamma_h$ becomes normally elliptic. Then it can no longer be a part of the NHIM. Thereby the NHIM looses its previous boundary. For energies just a little larger than this limit value there are still surviving normally hyperbolic KAM curves close to the orbit $\Gamma_h$ and the outermost one of them takes over the role of the boundary of the NHIM. For still larger energy these KAM curves disappear and then the orbits from the large chaotic sea can fall over the edge and disappear from the NHIM surface and the NHIM falls apart. A few fragments around the tangentially stable and normally hyperbolic fixed points survive. This is similar to the scenario found in the example of \citet{GJ15}. For $E = -2534$ also the two orbits split off from the orbit $\Gamma_v$ in its first pitchfork bifurcation become normally elliptic. Then also these two orbits are lost from the NHIM, i.e. the centres of the corresponding two islands in the restricted Poincar\'e map are lost. This is another step towards the total decay of the NHIM.

In Fig. \ref{maps} we observe that for higher energy the chosen projection into the $(\phi,L)$ plane is no longer appropriate to show well the islands around the tilted loop orbits coming from the first pitchfork bifurcation of the Lyapunov orbit $\Gamma_v$. Therefore we present in Fig. \ref{maps3} the same data sets again in another presentation, this time it is a perspective view in the $(x,y,p_y)$ space. Here we see very clearly one of the side islands and also the part of the large chaotic sea between the central main island and the side island. Of course, the other side island (the one at negative values of $y$) is obtained as mirror image from the one at positive values of $y$. In addition, the presentation in Fig. \ref{maps3} together with Fig. \ref{maps} illustrates how the NHIM surface becomes bowl shaped for energies above -3100.

The scenario presented in Fig. \ref{maps} shows how with increasing energy the effects of nonlinearity increase leading to more and more complicated structures in the dynamics restricted to the NHIM. As part of this scenario we observe how with increasing perturbation secondary island chains first gain importance and later decay into the large chaotic sea grown out of the main separatrix. This is the generic behaviour of 2-dimensional symplectic maps in analogy to the standard map \citep[see e.g.,][]{C79}.

So far we have concentrated the discussion on the NHIM over the index-1 saddle point $L_2$. Of course, because of symmetry reasons there is an equal one over the saddle point $L_3$ and one of these two NHIMs is transformed into the other one by a rotation around the $z$ axis by an angle $\pi$.

Now the reader may ask whether there are important subsets also over the index 2 saddle points $L_4$ and $L_5$. The answer is ``no" because of the following arguments: The only orbit which stays over the saddle point $L_4$ for ever is a periodic orbit oscillating up and down in $z$ direction. This orbit is hyperbolic in all its normal directions, i.e. in $x$ and in $y$ motion. Accordingly in the 4-dimensional Poincar\'e map it appears as a fixed point which is hyperbolic in all 4 directions. If we like we can interpret this hyperbolic fixed point as a 0-dimensional NHIM in the map. In the map it has a 2-dimensional stable manifold and a 2-dimensional unstable manifold. These stable and unstable manifolds can create homoclinic/heteroclinic intersections and can thereby imply some invariant set in the neighbourhood of the saddles $L_4$ and $L_5$. However, such surfaces of codimension 2 do not divide anything in a 4-dimensional embedding space. They are not able to direct and channel the general dynamics. The structures generated by these manifolds are too open and too low dimensional to be of great general importance.

\begin{figure*}
\centering
\resizebox{\hsize}{!}{\includegraphics{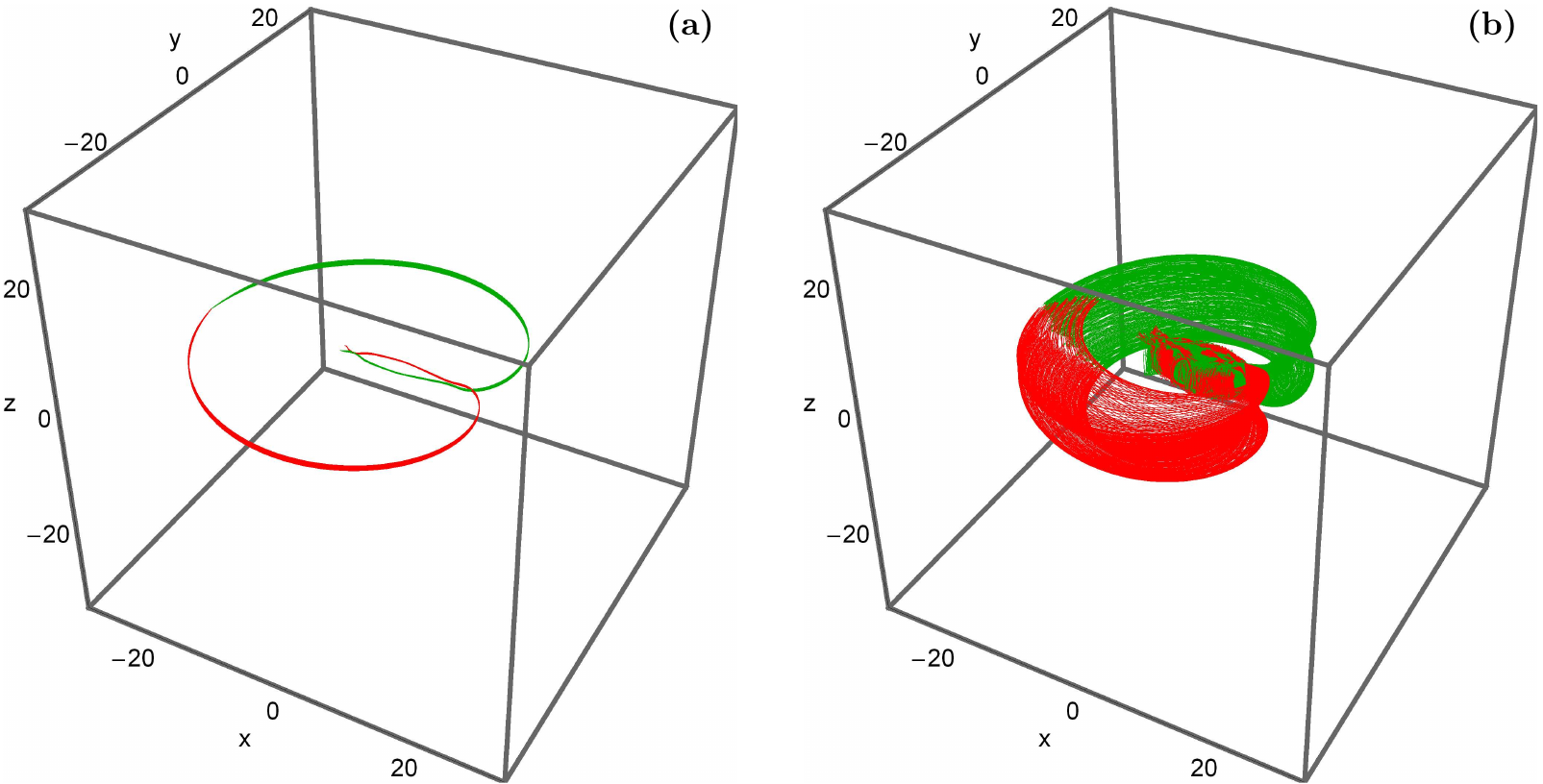}}
\caption{The stable manifold $W^s(\Gamma_{v/d})$ (green) and the unstable manifold $W^u(\Gamma_{v/d})$ (red), when (a-left): $E = -3240$ and (b-right): $E = E(L_4)$. (For the interpretation of references to colour in this figure caption and the corresponding text, the reader is referred to the electronic version of the article.)}
\label{mans}
\end{figure*}

\subsection{Stable and unstable manifolds of the Lyapunov orbits}
\label{sum}

For the next section on rings and spirals we need a good understanding of the stable and unstable manifolds of the NHIM, we will call them $W^s(S_E)$ and $W^u(S_E)$ respectively. As has become evident in the previous subsection \ref{pmap}, the two Lyapunov orbits $\Gamma_h$ and $\Gamma_v$ are the central elements of the NHIM. Therefore the stable and unstable manifolds of these prominent periodic orbits (we will call them $W^s(\Gamma_h)$, $W^u(\Gamma_h)$, $W^s(\Gamma_v)$ and $W^u(\Gamma_v)$ respectively) play the key role of central structures in $W^s(S_E)$ and $W^u(S_E)$. So we begin to gain an understanding
of $W^s(S_E)$ and $W^u(S_E)$ by a description of $W^s(\Gamma_h)$, $W^u(\Gamma_h)$, $W^s(\Gamma_v)$ and $W^u(\Gamma_v)$. When we make statements about the Lyapunov orbit $\Gamma_v$ then these statements hold equally well for the orbit $\Gamma_d$ which is the continuation of the orbit $\Gamma_v$ for higher energies, frequently we will write $\Gamma_{v/d}$ to point out that we talk about these two orbits simultaneously.

Imagine a single energy shell in the phase space for fixed energy $E$, it is a 5-dimensional manifold which we call $P_E$. A periodic orbit for fixed energy $E$ is a closed curve in $P_E$. The NHIM surface $S_E$ is a 3-dimensional sub-manifold of $P_E$ and the particular periodic orbits $\Gamma_h$ and $\Gamma_{v/d}$ lie in $S_E$. In the whole energy interval in which we are interested in this article (it is the energy interval $(E(L_2),E(L_4)))$ these orbits are normally unstable and with the exception of the very small energy interval between the two pitchfork bifurcations they are tangentially stable. Also the Lyapunov orbit $\Gamma_h$ is normally unstable and tangentially stable in this whole energy interval. Periodic orbits with these properties have stable and unstable directions normal to the surface $S_E$. Therefore these orbits have 2-dimensional stable and unstable manifolds $W^s(\Gamma_h)$, $W^u(\Gamma_h)$, $W^s(\Gamma_{v/d})$ and $W^u(\Gamma_{v/d})$ in the energy shell $P_E$.

First some remarks to the Lyapunov orbit $\Gamma_h$: This periodic orbit is the only orbit of $S_E$ lying in the plane $S_z$ defined by $z = 0, p_z = 0$, it is the intersection between $S_E$ and $S_z$. Also the unstable directions of the orbit $\Gamma_h$ lie in $S_z$. Therefore $W^s(\Gamma_h)$ and $W^u(\Gamma_h)$ are the intersections of $W^s(S_E)$ and $W^u(S_E)$ with $S_z$. Accordingly, we obtain $W^s(\Gamma_h)$ and $W^u(\Gamma_h)$ by a construction of the stable and unstable manifolds of the orbit $\Gamma_h$ under the reduced 2-dof dynamics in $S_z$. The projection of the local segments of these surfaces into the position space have been plotted in Fig. 13(a-b) of Paper I for an energy value $E = -3239.5294$, which is just a little above the escape energy $E(L_2)$. The plot is obtained numerically by just running, for a finite time interval, many orbits with initial conditions close to the orbit $\Gamma_h$ under the influence of the reduced dynamics. Forward in time these orbits converge to $W^u(\Gamma_h)$ and backward in time they converge to $W^s(\Gamma_h)$. This figure remains valid for the 3-dof dynamics. The most important observation from this figure is that $W^s(\Gamma_h)$ and $W^u(\Gamma_h)$ form tubes which channel and guide the general flow over the saddle point of the effective potential.

Corresponding plots for $W^s(\Gamma_{v/d})$ and $W^u(\Gamma_{v/d})$ are a little more tricky because they are not lying in some lower dimensional coordinate plane. Numerically we have done the following. We have again chosen many initial conditions close to orbits $\Gamma_{v/d}$ and run the orbits for a finite time interval. Of course, forward in time these orbits do not converge exactly to $W^u(\Gamma_{v/d})$, they oscillate around $W^u(\Gamma_{v/d})$ and backward in time they do not converge exactly to $W^s(\Gamma_{v/d})$, they oscillate around $W^s(\Gamma_{v/d})$. In the energy shell $P_E$ the collection of these orbits traces out a 4-dimensional layer which consists of parts of $W^s(S_E)$ and $W^u(S_E)$. However, when this layer is very thin in the directions inside $W^s(S_E)$ and $W^u(S_E)$ but transverse to $W^s(\Gamma_{v/d})$ and $W^u(\Gamma_{v/d})$ then numerically it appears as if it would consist of the 2-dimensional surfaces $W^s(\Gamma_{v/d})$ and $W^u(\Gamma_{v/d})$. This still holds after a projection into the position space. This is how Fig. \ref{mans}(a-b) has been constructed. Panel (a) is for $E = -3240$ which is just a little above the saddle energy $E(L_2)$ and panel (b) is for $E = E(L_4)$. The time interval used for the cut off is $t \in [0, \pm 1.83]$\footnote{Remember that for the stable manifold we must integrate the initial conditions of orbits into the past.} for panel (a) and $t \in [0, \pm 2.78]$ for panel (b). Stable manifolds are plotted green and unstable manifolds are plotted red. Also the local segments of these manifolds form tubes which channel and direct the flow over the saddle. Note that the vertical Lyapunov periodic orbits are not visible in Fig. \ref{mans}(a-b) because they are located in the vicinity of the intersections of the stable and the unstable manifolds.

An important observation is that also the projections of the local segments of $W^s(\Gamma_{v/d})$ and $W^u(\Gamma_{v/d})$ remain close to the plane $z = 0$ even though $\Gamma_{v/d}$ is the orbit in $S_E$ which deviated most from $S_z$. Now remember the argumentation in subsection \ref{quad} that general orbits in $S_E$ do something very similar to a quasi-periodic superposition of orbits $\Gamma_{v/d}$ and $\Gamma_h$. This suggests that the general orbit in $W^s(S_E)$ or $W^u(S_E)$ moves between $W^s(\Gamma_h)$ and $W^s(\Gamma_{v/d})$ or between $W^u(\Gamma_h)$ and $W^u(\Gamma_{v/d})$, respectively. This suggests that the whole manifolds $W^s(S_E)$ and $W^u(S_E)$ are confined to a small layer around $S_z$ in the same way as $W^s(\Gamma_{v/d})$ and $W^u(\Gamma_{v/d})$ are confined around $S_z$. In addition, we can conclude that also $W^s(S_E)$ and $W^u(S_E)$ form tubes in $P_E$ which channel and direct the general flow over the saddle. Of course, their projection into the position space does no longer look like the projection of a hollow 2-dimensional tube, it is the projection of a 4-dimensional tube. But, most important, also these projections indicate the position space regions which are influenced by the flow over the saddle guided by $W^s(S_E)$ and $W^u(S_E)$. These arguments will be essential for the next section.

\subsection{The NHIMs as the source of the global chaos}
\label{cha}

So far we have only studied the local segments of the stable and unstable manifolds of the NHIMs. Now a few remarks on their global behaviour and their global implications. On large scale these manifolds grow folds and tendrils on an infinity of levels of hierarchy and thereby create fractal structures. Now imagine this infinity of structure in the inner potential region. We have a stable and also an unstable manifold with all its folds and tendrils coming from both sides, i.e. one from the neighbourhood of the saddle $L_2$ and also one from the neighbourhood of the saddle $L_3$. Then it is understandable that mutual intersections between stable and unstable manifolds can not be avoided. Intersections between stable and stable manifolds or unstable and unstable manifolds are forbidden because any point in phase space has a unique past and a unique future orbit. But transverse intersections between stable and unstable manifolds are the rule in generic Hamiltonian systems. If such intersections occur between stable and unstable manifolds from the same NHIM then they are called homoclinic. If they occur between stable and unstable manifolds from different NHIMs then they are called heteroclinic. Iterated images of homoclinic/heteroclinic intersections must be further homoclinic/heteroclinic intersections because the stable and unstable manifolds are invariant subsets. Therefore any transverse intersection implies an infinity of further transverse intersections and implies an infinite fractal tangle between the involved stable and unstable manifolds. In addition, any tendril of some level of the hierarchy is the accumulation set of an infinity of further tendrils of higher levels of the hierarchy. This tangle is a higher dimensional generalisation of the well known horseshoes from 2-dimensional Poincar\'e maps. For general information on higher dimensional horseshoes see sections 2.3 and 2.4 in \citet{W88}.

In the neighbourhood of the homoclinic/heteroclinic intersections there exists an countable infinity of periodic orbits, among them are orbits with periods above any limit. In particular, near the heteroclinic intersections we find periodic orbits oscillating between the neighbourhoods of $L_2$ and of $L_3$. In this respect these orbits are very similar to the x1 orbits. However, there is one important difference. For energies above the saddle energy and for large values of the length $a$ of the bar these orbits make rather wide movements also in the $y$ direction, i.e. these orbits are not confined to the interior of the bar. Therefore they are no real x1 orbits, compare the discussion in section 3.2 of Paper I. In addition, we find near the homoclinic/heteroclinic intersections an over-countable infinity of truly chaotic bound (non-escaping) orbits.

For general initial conditions in the interior of the potential and outside of the regions of bounded regular orbits the following behaviour is generic: The initial condition does not lie exactly on any stable or unstable manifold, nor does it lie exactly on any periodic orbit or any bound chaotic orbit. But it lies in the neighbourhood of some tendril of a stable manifold. Accordingly this orbit moves along this stable manifold and after some time comes close to the corresponding NHIM. Then it depends on which side of the local segment of the stable manifold the orbit arrives. Coming on one side it passes the saddle and escapes. Coming on the other side it returns to the interior potential region. Of course, if it returns then later it will come close to some saddle region again, has again to make the decision to escape or to return depending on which side of the local segment of the stable manifold it arrives, etc. The stable manifolds of the invariant sets over the saddles are the division surfaces between qualitatively different behaviour. Here, it is essential that these stable manifolds are of codimension 1, since only surfaces of codimension 1 can divide the embedding space into distinct sides.

Any orbit which has come close to the NHIM for a while and later leaves the neighbourhood of the saddle, i.e. of the NHIM,
does this along and close to some branch of the unstable manifold of this NHIM, either the inner one or the outer one. Thereby it is clear that any large collection of orbits which go over the saddle with an energy close to the saddle energy trace out the position of the unstable manifolds of the NHIM.  This picture we must keep well in mind to understand the basic idea of section \ref{rs}.

When we have some general 2-dimensional surface $S$ of initial conditions in the phase space, then this surface is intersected by the fractal bundle of the stable manifolds. Since these stable manifolds are of codimension 1 in the embedding space, the intersections of the various segments of the stable manifold with $S$ are 1-dimensional curves. And if we have a fractal bundle of segments of the stable manifold then the corresponding intersections consist of a fractal of intersection curves. And the fractality of this collection of intersection curves coincides with the fractality of the homoclinic/heteroclinic tangle created by the NHIMs. Because the fate of general orbits depends on which side of the stable manifold they start and whether they start in a tendril of the stable manifold coming from $L_2$ or in a tendril of the one coming from $L_3$, it is understandable that the stable manifolds are the basin boundaries of the escape basins. This explains how the fractal basin structure observed numerically in section \ref{orbdyn} is generated by the saddle NHIMs and their stable manifolds.

All these considerations show that stable manifolds of codimension 1 play an essential role in the dynamics. Only normally hyperbolic invariant subsets of codimension 2 can have stable and unstable manifolds of codimension 1. And finally, only over index-1 saddles of the effective potential we have NHIMs of codimension 2. This chain of arguments taken together explains why exactly the NHIMs of codimension 2 over the index-1 saddles generate the global chaos properties of open systems and direct and channel the escape.

\subsection{Some comments on our NHIM development scenario}
\label{com}

In our example the NHIM plays two important roles. First and as seen in Figs. \ref{maps} and \ref{maps3}, a 2-dimensional NHIM is the ideal domain to present the perturbation scenario in form of 2-dimensional plots. This can be done because the NHIM is invariant and therefore we have naturally the restriction of the general Poincar\'e map to this 2-dimensional subset. If this presentation of the scenario would be the only purpose than also any other 2-dimensional invariant surface would do the job, for example a surface which is invariant because of reasons of discrete symmetry. For an example of the study of the restricted Poincar\'e map on this type of invariant domain \citep[see e.g.,][]{LRJ15}. In this sense an invariant 2-dimensional surface in the domain of the map can serve as a screen to display the development scenario of the system.

In our model for the barred galaxy a good example for a subset invariant because of reasons of discrete symmetry is the 4-dimensional subset $S_z$ of the phase space with $z = 0$ and $p_z = 0$. The restriction of the dynamics to this subset is exactly the reduced 2-dof system investigated in Paper I. There we have shown in Fig. 4 some 2-dimensional Poincar\'e plots of the reduced system which helped a lot to understand the reduced dynamics. This reduction is possible because of the invariance of the Hamiltonian from Eq. (\ref{ham}) under the symmetry $ z \rightarrow -z, p_z \rightarrow - p_z$. Our effective potential is increasing in $z$ direction, accordingly in the quadratic approximation the $z$ motion is oscillatory, see Eq. (\ref{zosc}). Also after the inclusion of small perturbations by higher order terms of the potential the $z$ motion remains mainly oscillatory. This means that the surface $S_z$ is not hyperbolic in normal direction, i.e. $S_z$ is not a NHIM, and it does not have global stable and unstable manifolds of codimension 1 in the 6-dimensional phase space of the full 3-dof system. Therefore we can not claim that it directs and channels the dynamics of the full 3-dof system. And in this sense the dynamics restricted to $S_z$ is not representative for the full 3-dof dynamics. These arguments may provide some motivation for the considerations of the next paragraph.

However, we are interested not only in the dynamics on a particular lower dimensional surface, we are interested to gain information on the global full dimensional dynamics. And then it is important whether or not the dynamics on the particular invariant surface is essential and representative for the global dynamics. Here the NHIM properties enter. If the invariant surface is a true NHIM then its restricted dynamics has global implications. A NHIM has stable and unstable manifolds of codimension 1. Therefore these manifolds divide the phase space, form tubes and channels and direct the global flow to a large extent. We can imagine that the dynamical structures on the NHIM itself are transported along the stable and unstable manifolds to far away regions of the phase space and thereby determine important dynamical structures also in such far away regions. In this sense the codimension 2 NHIMs are the most important elements of the skeleton of the global dynamics. The implications of this global influence of the NHIMs in our particular example of the barred galaxy will be elaborated in the section on rings and spirals. Note that to represent the perturbation scenario on a 2-dimensional domain we need a NHIM of dimension 2 in the full domain of the map. On the other hand to have a subset which directs the global flow we need a NHIM of codimension 2. Only for 3-dof systems a NHIM can fulfil both of these conditions at the same time. This is, only in 3-dof systems we can use the same NHIM for the two different purposes.

Finally we can ask whether the scenario found here is typical or not for any systems having saddle points in the form of Lagrange points $L_2$ and $L_3$ in the effective potential in the rotating system. There is the interesting example of the restricted three-body problem. In \citet{JM99} the restriction of the full 3-dof dynamics to the centre manifold of the saddle point and the related restricted Poincar\'e map have been constructed. Restrictions to a NHIM and restrictions to a centre manifold are the same basic idea and it is interesting to compare our restricted map with the corresponding plots presented in \citet{JM99} (see for example figures 3 and 6 in this publication). Also the authors use the intersection condition $z = 0$, therefore also in their case the central fixed point of the map represents the Lyapunov orbit $\Gamma_v$ and the energetic boundary represents the Lyapunov orbit $\Gamma_h$. They find in their example one great difference in comparison to our system. In their system only 2 stable tilted loop orbits split off from the orbit $\Gamma_h$ which after this bifurcation remains tangentially unstable. Furthermore, in their system the orbit $\Gamma_v$ always remains stable and does not split off tilted loop orbits. At the moment it is not clear to us what is the cause of this qualitatively different behaviour in the two systems. In this context it is interesting to have another look at the important periodic orbits in our system. When the tilted loop orbits split off from the vertical Lyapunov orbit, then they start as orbits very similar to their parent orbit, i.e. to the vertical Lyapunov orbit. However, for increasing energy they rapidly change their shape and become a lot more similar to the horizontal Lyapunov orbit (see again Fig. \ref{orbs}). Some additional insight into the relation between the Lyapunov orbits and the tilted loop orbits might be gained by an investigation of the development scenarios in the barred galaxy model for all kinds of parameter changes.

Some preliminary investigations of a change of $a$ and $\Omega_{\rm b}$ suggest the following: For smaller values of $a$ or of $\Omega_{\rm b}$ (i.e. when the effect of the bar becomes smaller) the replacement of the original Lyapunov orbit $\Gamma_v$ by the orbit $\Gamma_d$ does not happen. The Orbit $\Gamma_v$ continues to high values of the energy without the appearance of the two saddle-centre bifurcations. Therefore, when we are free to change other parameters as well during the change of the energy, then we can connect the orbit $\Gamma_v$ smoothly with the orbit $\Gamma_d$ without running through any bifurcations. This shows that in reality the orbit $\Gamma_d$ can be considered the direct continuation of the orbit $\Gamma_v$. We can imagine that in a bifurcation diagram analogous to Fig. \ref{fpo}a but over the 2-dimensional $(E,a)$ plane or over the 2-dimensional $(E,\Omega_{\rm b})$ plane the surface formed by the orbits $\Gamma_v$, $\Gamma_c$ and $\Gamma_d$ is the surface of a cusp catastrophe. For the geometry of this surface the reader can find more information in section 9.3 in \citet{PS78}. Otherwise the change of $a$ or of $\Omega_{\rm b}$ does not cause any qualitative change of the scenario of the Lyapunov orbits. Accordingly the differences between our scenario and the one of \citet{JM99} must have causes different from the properties of the bar. Therefore this problem lies outside of the scope of the present article.

The bifurcations of NHIMs and the development scenarios of NHIMs are problems little explored up to now. Therefore we are not yet able to put the scenario seen in the present example into some well known general scheme. More information regarding observations of bifurcations of NHIMs can be found in \citet{AB12,MS14,MCE13,LST06,TTK11,TTK15,TTT15}. We hope that our observations contribute to the collection of information on this difficult but very interesting topic.

\section{Formation of rings and spirals}
\label{rs}

If there are stars in the interior region of the galaxy with an energy high above the threshold energy $E(L_2)$ then such stars will leave the interior region rather fast and the interior region will have lost such stars long time ago. Let us consider now stars with an energy below the threshold but close to it and moving in the inner part of the galaxy. Such stars have occasional interactions among themselves and with other objects and thereby their energy can be changed slightly and it may come a little above the threshold. Then such stars are exactly the ones for which the structure of the NHIMs over the Lagrange points $L_2$ and $L_3$ and their stable and unstable manifolds become highly relevant. First these stars can come close to the saddle points along the stable manifolds of the NHIMs, and then they have two possibilities for their further motion. First, from the neighbourhood of the saddle they can return to the inner region of the galaxy along the inner branches of the unstable manifolds of the NHIMs. Second, they can leave to the outer region of the galaxy along the outer branches of the unstable manifolds of the NHIMs. Which one of these two possibilities is realized depends on the finest details of the initial conditions of the stars, it depends on which side of the local branch of the stable manifold they start their orbits. Stars which return to the inner well of the effective potential will come back to the neighbourhood of the saddle points later and can then eventually escape later. Of course, the same possibilities exist for stars which come close to the saddles from the outside along the outer branches of the stable manifolds of the NHIMs. The stars which happen to leave to the outer part of the galaxy may cause perturbations in the disc in the outer region and trigger the formation of rings and spirals. Such stars only escape with very small rates and over very long times and are supposed to maintain the ring and spiral structure over billions of years.

\begin{figure*}
\centering
\resizebox{\hsize}{!}{\includegraphics{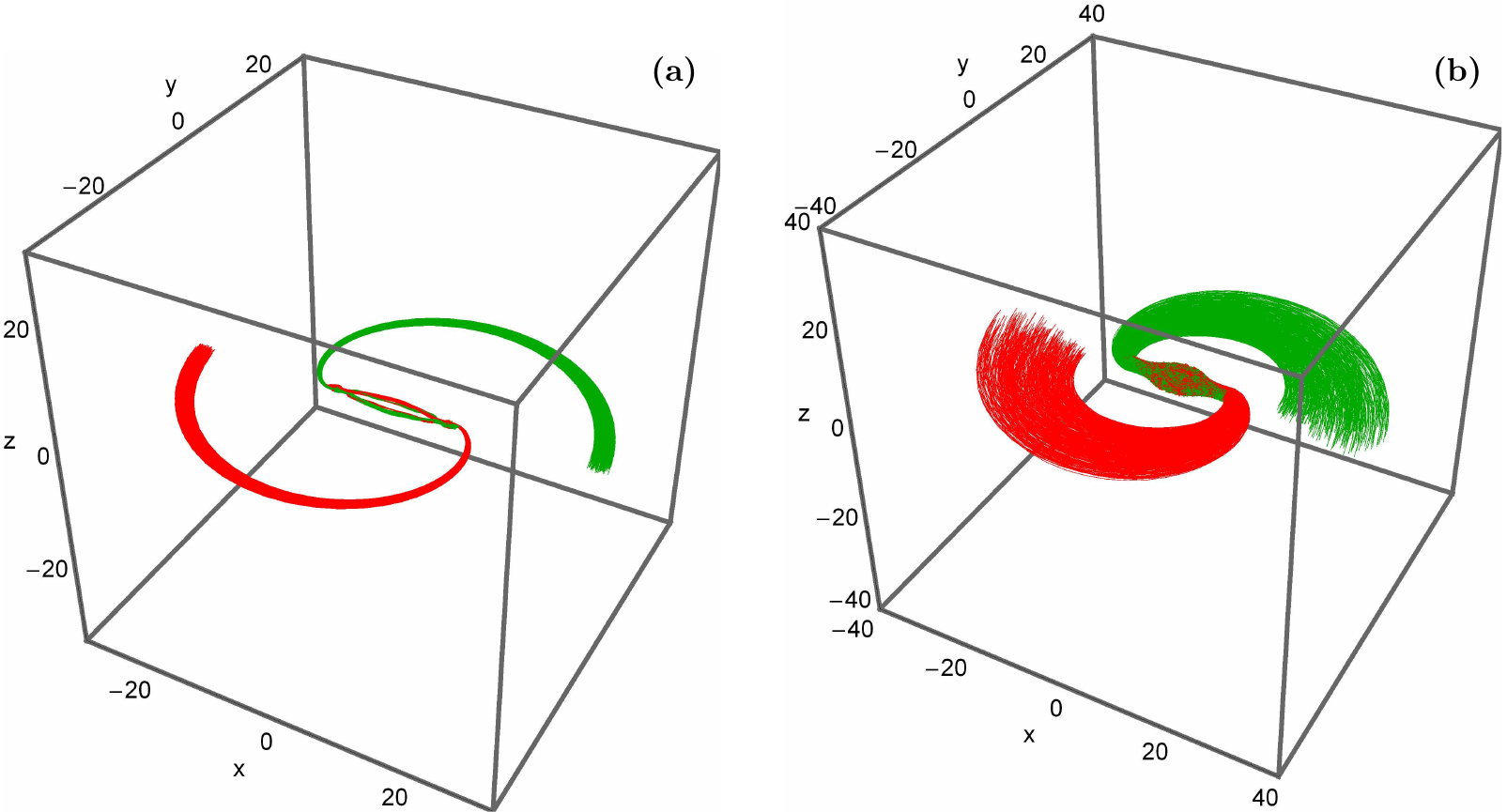}}
\caption{The projection of the local segments of $W^u(S_E)$ into the $(x,y,z)$ position space, when (a-left): $E = -3240$, and (b-right): $E = E(L4)$. The manifolds from the NHIM over the saddle point $L_2$ are shown in red colour, while the ones from the symmetrically placed NHIM over $L_3$ are shown in green colour. (For the interpretation of references to colour in this figure caption and the corresponding text, the reader is referred to the electronic version of the article.)}
\label{nhim}
\end{figure*}

When an orbit starts in the neighbourhood of the saddle point $L_2$ then in forward direction (i.e. in the future) it converges automatically against $W^u(S_E)$. This observation provides the idea for a numerical construction of $W^u(S_E)$. We randomly select 1000 initial conditions close to $L_2$ all with the same energy and we let the orbits run for a finite time. In particular these initial conditions are taken from the corresponding maps presented in Figs. \ref{maps} and \ref{maps3} (note that all these 3-dimensional orbits have $z_0 = 0$). The whole collection gives a good representation of the unstable manifold. However, before plotting $W^u(S_E)$ we have to consider 3 minor problems. First, it is too difficult to produce plots giving a good impression how $W^u(S_E)$ is located in the 5-dimensional energy shell $P_E$. In addition, our further discussion in this section will focus on the position space. These two points are taken care of by a projection of $W^u(S_E)$ into the position space. Third, the complete surface $W^u(S_E)$ has an infinite extension and has an infinity of folds. For our further discussions only the local segments (i.e. the parts emanating directly from the NHIM) are of importance. Therefore, we have to cut off the surface appropriately, we restrict the surface to its local segments by following the orbits mentioned above over a finite time interval only. With all these considerations in mind we plot in Fig. \ref{nhim}(a-b) the projection of the local segments of $W^u(S_E)$ into the position space. Panel (a) is for $E = -3240$, while panel (b) is for $E = E(L4)$. The time interval used for the cut off is $t \in [0, 2.85]$ for panel (a) and $t \in [0, 4.65]$ for panel (b). We include the manifolds from the NHIM over $L_2$ in red colour and the ones from the symmetrically placed NHIM over $L_3$ in green colour. The inner as well as the outer local segments are included.

The important observations are: For the energy close to $E(L_2)$ the unstable manifold fills a rather fine tube only with a very sharp boundary and it is restricted to a very thin layer around the surface $z = 0$. With increasing energy however, this tube slowly becomes wider but still remains confined close to $z = 0$. Only when the energy approaches $E(L_4)$, then $W^u(S_E)$ becomes wide and fuzzier. However, our further discussions below will concentrate on an energy region close to $E(L_2)$. Then the main conclusion is, that in the relevant energy interval the local segments of $W^u(S_E)$ affect a well delimited part of the position space only. And as Fig. \ref{nhim} suggests, this delimited region is related to the spiral structure of the galaxy.

\begin{figure*}
\centering
\resizebox{0.9\hsize}{!}{\includegraphics{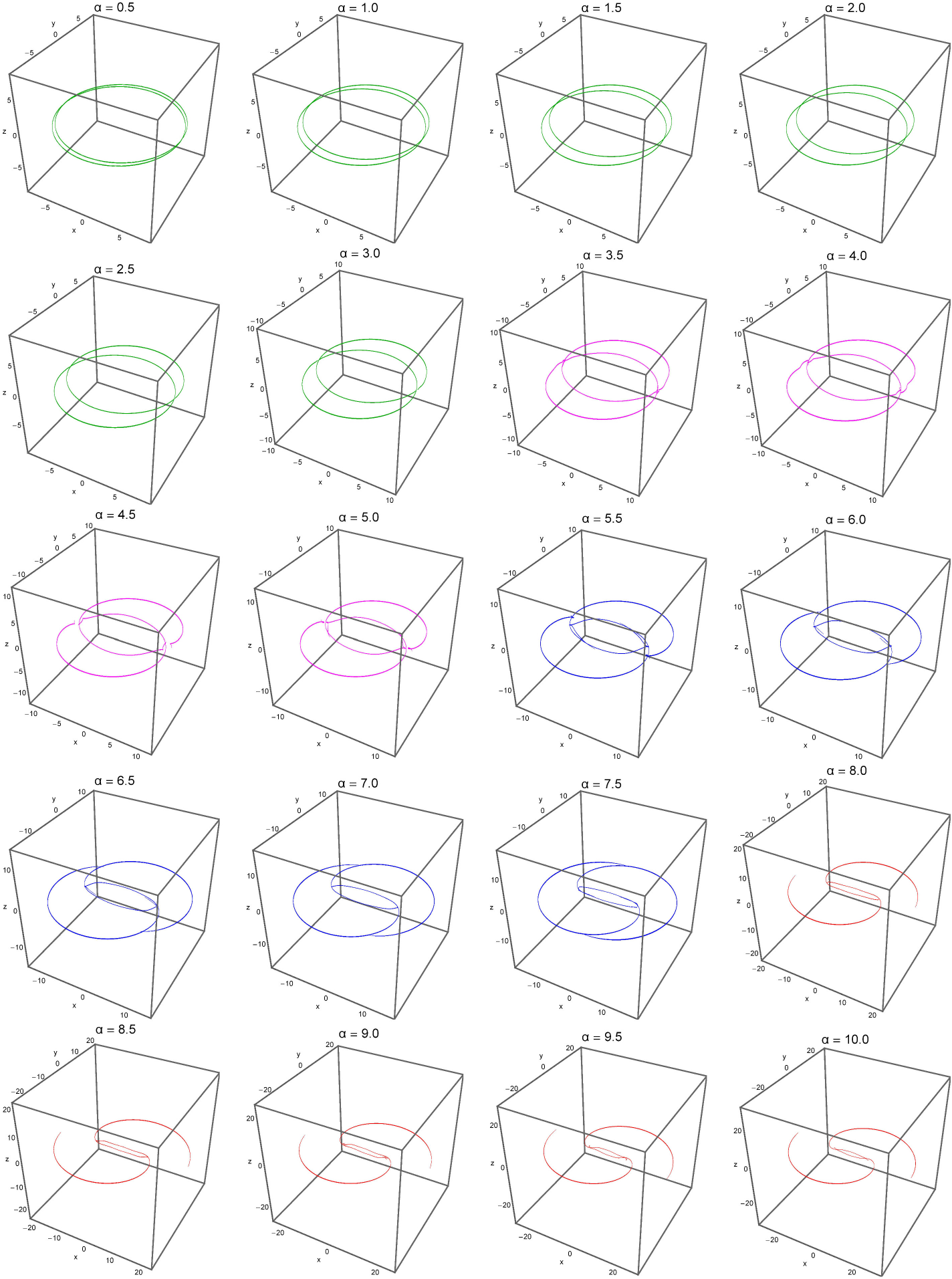}}
\caption{Morphologies of the unstable manifolds $W^u(\Gamma_{v_2})$ and $W^u(\Gamma_{v_3})$ for several values of the semi-major axis $a$ of the galactic bar. For all models we have $\widehat{C} = 0.0001$. The manifolds are plotted in a different colour which is determined by the corresponding morphology. The colour code is as follows: $R_1$ rings (green); $R_1'$ pseudo-rings (magenta); $R_1R_2$ rings (blue); open spirals (red). (For the interpretation of references to colour in this figure caption and the corresponding text, the reader is referred to the electronic version of the article.)}
\label{rgsp}
\end{figure*}

In Paper I (see Fig. 14) we proved that in the 2-dof system the value of the semi-major axis of the bar $(a)$ strongly influences the structure of the unstable manifolds $W^u(LO_2)$ and $W^u(LO_3)$, where $LO_2$ and $LO_3$ are the horizontal Lyapunov periodic orbits in the vicinity of the Lagrange points $L_2$ and $L_3$, respectively. It would be very interesting to investigate how the value of $a$ affects the structure of the unstable manifolds in the 3-dof system. The local segments of $W^u(\Gamma_{v_2})$ and $W^u(\Gamma_{v_3})$ for values of $a$ from 0.5 up to 10 in steps of 0.5 are presented in Fig. \ref{rgsp}. $\Gamma_{v_2}$ and $\Gamma_{v_3}$ are the vertical Lyapunov periodic orbits near the Lagrange points $L_2$ and $L_3$, respectively. For every value of $a$ the energy level $E$ is chosen such that it is $\widehat{C} = 0.0001$\footnote{The energy of escape $E(L_2)$ can be used in order to define a dimensionless energy parameter as $\widehat{C} = \left(E(L_2) - E\right)/E(L_2)$, where $E$ is some other value of the energy integral. This dimensionless energy parameter $\widehat{C}$ makes more convenient the reference to energy levels above the escape energy.} above the respective saddle energy which also depends on the semi-major axis of the bar. We would like to stress that the existence of the unstable manifolds $W^u(\Gamma_{v_2})$ and $W^u(\Gamma_{v_3})$ is only a necessary but by no means a sufficient condition for the corresponding stellar structure to develop. This is true if we think of the following argument: In theory a manifold may be present (obtained by the numerical integration) for a particular galaxy model. In a real barred galaxy however, with similar dynamical properties (like those taken into account in the corresponding mathematical model) the manifold may not be able to trap inside of it a sufficient amount of stars and therefore the corresponding stellar structure will not be observable.

\begin{figure*}
\centering
\resizebox{0.9\hsize}{!}{\includegraphics{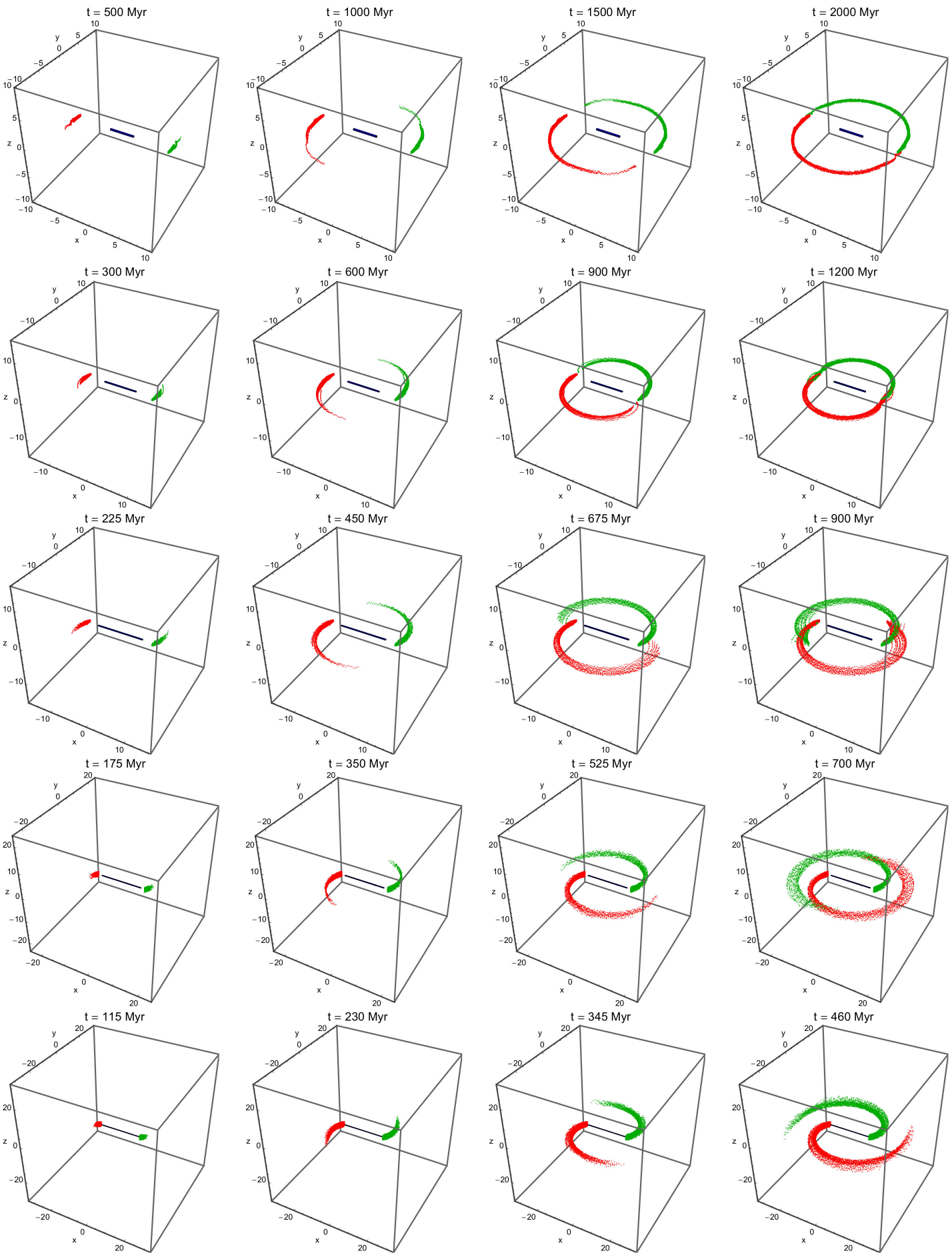}}
\caption{The distribution of the position of stars in the configuration $(x,y,z)$ space initiated $(t = 0)$ within the Lagrange radius, for $\widehat{C} = 0.001$ and $\Omega_{\rm b} = -4.5$. The green segment contains stars that escaped through $L_2$, while the red segment contains stars that escaped through $L_3$. The horizontal black bar in the interior region indicates the total length of the bar. (first row): $a = 2$; (second row): $a = 4$; (third row): $a = 6$; (fourth row): $a = 7.5$; (fifth row): $a = 10$. (For the interpretation of references to colour in this figure caption and the corresponding text, the reader is referred to the electronic version of the article.)}
\label{ssg}
\end{figure*}

Looking at the development scenario presented in Fig. \ref{rgsp} we may conclude that:
\begin{itemize}
  \item For relatively small values of the semi-major axis of the bar, where $a \in [0.5,3.5)$, the local segments of the unstable manifolds trace out a ring structure around the interior region of the galaxy. The structure where the major axis of the ring points into the $y$ direction is called $R_1$ ring. In this case, the unstable manifolds from one side come very close to the opposite saddle point thus forming approximate heteroclinic separatrix connections.
  \item When $a \in [3.5,5.5)$ the heteroclinic connections are clearly broken. Now the major axis of the ring rotates in negative orientation with increasing value of the semi-major axis of the bar. These structures are called $R_1'$ pseudo-rings.
  \item When $a \in [5.5,7.5]$ the unstable manifolds coming from one side connect to the unstable manifold from the other side in a point far away from the saddle. These structures are called $R_1R_2$ ring. It is seen that the major axis of the $R_1R_2$ ring still rotates in negative orientation with increasing value of $a$.
  \item When $a > 7.5$ the orientation of the major axis of the $R_1R_2$ rings approaches the $x$ axis, then the rings break and the unstable manifolds form twin open spirals which begin very close to the two ends of the bar.
\end{itemize}
We would like to point out that the classification of the stellar structures shown in Fig. \ref{rgsp} has been preformed by eye inspection following the usual method of observational astronomy according to which real galaxies are classified. The reader can find more useful information regarding stellar structures in \citet{BC96}.

We observe that the types of the stellar structures (rings or spirals) for all the examined values of $a$ completely coincide with corresponding ones obtained in Paper I from the 2-dof system. Furthermore, as in the 2-dof system, we did not find persistent rings of type $R_2$ (rings with the major axis pointing into the bar direction). Of course, the value of the major-semi axis of the bar is not the only dynamical quantity that influences the geometry of the stellar structures. In previous works the influence of other dynamical parameters of the bar (i.e., the axial ratio, the Lagrange radius related with the angular velocity of the bar, the central density, etc) has been studied in a variety of galactic potentials \citep[see e.g.,][]{ARGM09,ARGBM09,RGAM07}. This extensive investigation of all the involved parameters of the bar is out of the scope of the present paper.

In the 2-dof system we proved that our new barred galaxy model can realistically describe the formation as well as the time-evolution of rings and spirals. Now we shall examine if the full 3-dof system retains this ability. As in Paper I, the only variable parameter will be the semi-major axis of the bar varying in the interval $(a \in [0.5, 10])$, while the values of all the other parameters remain constant according to the standard model. Usually when performing galactic simulations the bar rotates counter-clockwise (in direct sense with respect to the rotation of the galaxy itself). For this reason, the sign of the angular velocity of the bar should change $(\Omega_{\rm b} = -4.5)$.

For the initial condition of the orbits we define a dense uniform 3-dimensional grid of size $N_x \times N_y \times N_z = 100 \times 100 \times 100$, with $p_{x_0} = p_{z_0} = 0$, while both signs of $p_{y_0}$ (obtained through the Jacobi integral of motion) are allowed. All the initial conditions of the orbits lie in the interior region of the galaxy inside the Lagrange radius, with $\widehat{C} = 0.001$. We numerically integrate the initial conditions of the 3-dimensional orbits and we record the output of all orbits. This allow us to monitor the formation as well as the time-evolution of the stellar structures constructed by the stars that escape through $L_2$ and $L_3$. The time-evolution of the $(x,y,z)$ position of stars for five values of the semi-major axis of the bar is illustrated in Fig. \ref{ssg}. The density of the points along a star orbit is taken to be proportional to the velocity of the star, according to Paper I. This means that a point is plotted (showing the exact position of a star on the configuration $(x,y,z)$ space), only if an integer counter variable which is increased by one at every step of the numerical integration, exceeds the corresponding velocity of the star. Adopting this numerical approach we can partially replicate a real $N$-body simulation of a barred galaxy, where the density of the stars will be highest where the corresponding velocity is lowest.

We observe in Fig. \ref{ssg} that initially the vast majority of the stars remain inside the interior region of the galaxy. As time goes by however, stars start to escape through the saddle points $L_2$ (green) and $L_3$ (red) thus indicating the formation of stellar structures. These structures grow in size with increasing time and the final morphologies are fully revealed. Once more, the morphology of the final stellar structure strongly depends on the particular value of the semi-major axis of the bar. In particular, when $a = 2$ a $R_1$ ring is formed, when $a = 4$ a $R_1'$ pseudo-ring is present, for $a = 6$ and 7.5 we have the scenario of a $R_1R_2$ ring, while for $a = 10$ a pair of twin spiral arms is developed. Therefore we conclude that the final stellar structures in all five models coincide not only with that derived earlier from the unstable manifolds (see Fig. \ref{rgsp}) but also with the corresponding ones from the 2-dof system investigated in Paper I (see Fig. 15). Thus taking into account and combining the numerical results given in Figs. \ref{rgsp} and \ref{ssg} one may reasonably conclude that the morphologies of the final stellar structures (that is rings or spirals) (i) depend on the value of the semi-major axis of the bar and (ii) they are completely unrelated with the particular distribution of the initial conditions of the orbits (near the unstable Lyapunov orbits or uniformly spread across all over the interior region). Additional numerical simulations (not shown here for saving space) suggest that similar stellar structures are developed for other (lower or higher) values of Jacobi integral of motion.

\begin{table}
\begin{center}
   \caption{The value of the semi-major axis $a$, the morphological type of several galaxies and the corresponding prediction of our dynamical model regarding the type of the observed stellar structures (rings or spirals).}
   \label{table1}
   \setlength{\tabcolsep}{6.0pt}
   \begin{tabular}{@{}lccc}
      \hline
      Galaxy & Type & $a$ (kpc) & Model prediction \\
      \hline
      NGC 1326 & $(\rm R_1)$SAB(r)0/a      &  3.1 & $\rm R_1$ ring \\
      NGC 3504 & $(\rm R'_1)$SAB(rs)ab     &  3.7 & $\rm R'_1$ ring \\
      IC 4214  & $(\rm R'_1)$SAB(r)a       &  4.4 & $\rm R'_1$ ring \\
      NGC 5248 & $(\rm R'_1)$SAB(rs)bc     &  4.6 & $\rm R'_1$ ring \\
      NGC 7552 & $(\rm R'_1)$SB(s)ab       &  5.0 & $\rm R'_1$ ring \\
      NGC 1672 & $(\rm R'_1)$SB(r)bc       &  5.2 & $\rm R'_1$ ring \\
      NGC 3081 & ($\rm R_1 R'_2$)SAB(r)0/a &  5.7 & $\rm R_1R'_2$ ring \\
      NGC 6782 & ($\rm R_1 R'_2$)SB(r)a    &  6.4 & $\rm R_1R'_2$ ring \\
      NGC 1241 & SAB(rs)b                  &  7.8 & spirals \\
      NGC 1819 & SB0                       &  8.3 & spirals \\
      NGC 5020 & SAB(rs)bc                 &  8.6 & spirals \\
      IC 4933  & SB(rs)bc                  &  8.8 & spirals \\
      NGC 5905 & SB(rs)bc                  &  9.4 & spirals \\
      NGC 5135 & SB(1)ab                   &  9.5 & spirals \\
      NGC 1343 & SAB(s)b                   &  9.7 & spirals \\
      NGC 1300 & SB(s)bc                   & 10.0 & spirals \\
      NGC 7771 & SB(s)a pec                & 10.3 & spirals \\
      NGC 7570 & SBa                       & 11.2 & spirals \\
      NGC 2595 & SAB(rs)c                  & 11.8 & spirals \\
      NGC 3313 & SB(r)b                    & 12.9 & spirals \\
      \hline
   \end{tabular}
\end{center}
\end{table}

It is evident from Fig. \ref{ssg} that with increasing value of the semi-major axis of the bar the time needed for final states of the morphologies to be developed decreases. Another interesting observation is the following: if we compare the time-evolution of the morphologies in the 2-dof system (see Fig. 15 in Paper I) and in the 3-dof system we see that for all models (values of $a$) the times in the 3-dof system are about three times larger than the corresponding ones of the 2-dof system. At the moment it is not yet clear to us why the time increases with increasing value of $a$. On the other hand, for the phenomenon according to which the times in the 3-dof system are about three times larger than those of the 2-dof system we may suggest the following explanation: In the 3-dof system the stars have one extra degree of freedom (with respect to the 2-dof system) thus they spend more time inside the interior region of the galaxy before they escape through the saddle points and start forming the different types of stellar structures.

\begin{figure}
\includegraphics[width=\hsize]{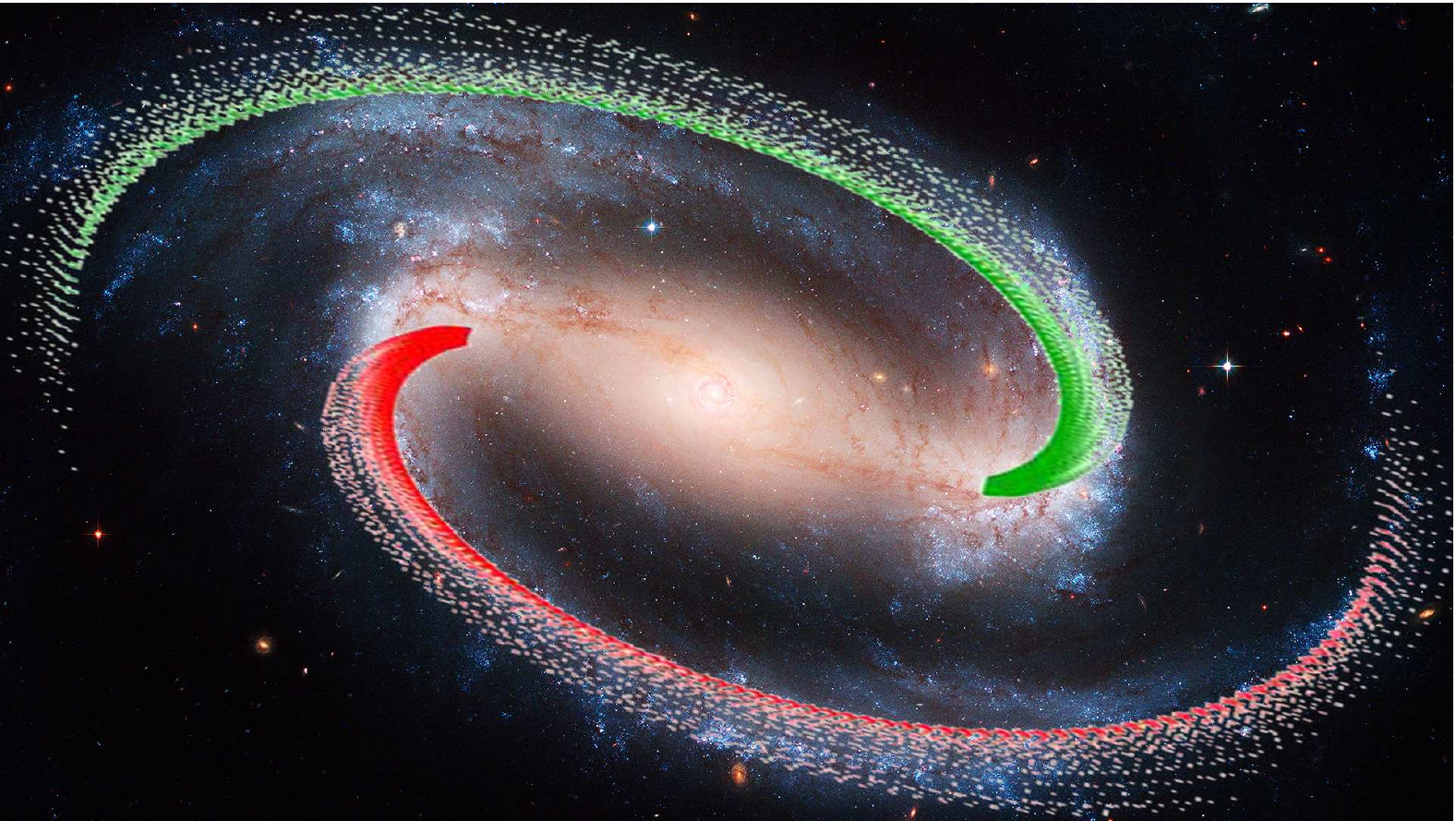}
\caption{A real image of the barred galaxy NGC 1300 where we have superposed the simulation output of our dynamical model at $t = 420$ Myr, when $E = -3225$. It is seen that the fit between the real and the corresponding simulated spiral arms is sufficient.}
\label{fit}
\end{figure}

Before closing this section we would like to connect the theoretical predictions of our dynamical model with corresponding data derived from observations. In Table \ref{table1} we provide for several galaxies the value of the semi-major axis $a$, as well as the morphological type as they were derived from the observational data presented in \cite{CKB10}. In the same table, the corresponding theoretical predictions of our model, regarding the final developed stellar structures (rings or spirals), are also given. We see that the predictions of our model coincide with the observational data. However, it should be noted that there are also other cases in which our model fails to predict the observed stellar formation. For instance in the galaxy NGC 5728 a $R_1$ ring has been observed even though the corresponding value of the semi-major axis is relatively high $(a = 12.7)$. In the same vein, for the galaxy NGC 1387, with $a = 2.2$, our model predicts an $R_1$ ring structure which is far from the observed open spirals. There are also examples, such as the NGC 5194 (also known as the Whirlpool galaxy), where the barred galaxies are in interaction with smaller satellite companion galaxies and therefore they are perturbed from the outside. Perturbations from the outside are certainly not contained in our dynamical model and therefore we should not expect to predict the stellar structures in such barred galaxies. In general terms, we may claim that our barred galaxy model can, in a way, predict the final stellar formations in a portion of isolated barred galaxies but by no means in all of them.

In Section \ref{galmod} we explained that the fiducial value of the semi-major axis $(a = 10)$ was chosen having in mind the barred galaxy NGC 1300. Following a similar procedure, as the one described earlier in Fig. \ref{ssg}, we conducted numerous numerical simulations, in a wide range of values of energy, in order to see whether or not our model can describe the spiral structure of this barred galaxy. Finally, for $E = -3225$ we obtained a sufficient match between the theoretical and the observational data. In Fig. \ref{fit} we have superposed the numerical simulation output at $t = 420$ Myr above the real image of the galaxy. One may observe that the twin spiral arms obtained from the numerical simulation almost coincide with the corresponding real ones.

\section{Conclusions}
\label{conc}

In this article we studied the 3-dof dynamics of a new simple and analytical model for barred galaxies. The total gravitational potential of the model is composed of four components: (i) a central spherically symmetric nucleus, (ii) a rotating bar, (iii) a flat disc and (iv) a spherically symmetric dark matter halo. This new dynamical model presents all the features observed in barred galaxies and expected for realistic models, while it has some clear advantages (i.e., its simplicity) over older models treated in the literature.

The most important elements of the skeleton of the dynamics are the NHIMs over the index 1 saddle points $L_2$ and $L_3$. Within these NHIMs the most important periodic orbits are the horizontal Lyapunov orbit, the vertical Lyapunov orbit and its
continuations and the tilted loop orbits split off from the vertical Lyapunov orbit. For these important periodic orbits we provide the bifurcation diagrams and detailed descriptions of their development scenario as function of the energy. The NHIMs and their development scenario are presented by the restriction of the Poincar\'e map to the NHIMs. The stable and unstable
manifolds of the NHIMs direct the flow over the saddle points and thereby are responsible to a large extent for the global structure formation of the galaxy, in particular for rings and spirals. Outside of the NHIMs we visualize the distribution of regular and chaotic motion by using colour-coded SALI plots. We study in detail the dependence of the structures on the semi-major axis of the bar.

We hope that the presented numerical outcomes shed some light on the role of the normally hyperbolic invariant manifolds in barred galaxies. In the third and last paper of the series (Paper III) we are going to investigate in detail the escape dynamics of the full 3-dof system. In particular, we shall conduct a thorough and systematic orbit classification in an attempt to locate the basins of escape towards the two channels of escape and to relate them with the corresponding escape times of the orbits.

\section*{Acknowledgments}

One of the authors (CJ) thanks DGAPA for financial support under grant number IG-100616. We would like to express our warmest thanks to the anonymous referee for the careful reading of the manuscript and for all the apt suggestions and comments which allowed us to improve both the quality and the clarity of our paper.

\section*{Appendix: Explanations for important concepts and terms from dynamical system theory}
\label{apx}

\begin{itemize}
  \item {\bf{Stable and unstable manifolds:}}
  Assume an invariant subset $S$ in the phase space $P$ of a dynamical system, which can be either a flow given by differential equations or an iterated map as for example the Poincar\'e map. Invariant means that any initial condition lying in $S$ leads to an orbit (past and future) also lying completely inside of $S$. Next assume that $S$ is unstable (hyperbolic) at least in some degrees of freedom. Then a general initial condition in the neighbourhood of $S$ leads to an orbit which runs away from $S$ exponentially. Most initial conditions do this in the past as well as in the future. However, there is an exceptional sub-manifold $W^s(S)$ such that initial conditions on $W^s(S)$ lead to orbits which converge toward $S$ in the future. And there is an exceptional sub-manifold $W^u(S)$ consisting of initial conditions leading to orbits which converge toward $S$ in the past. $W^s(S)$ is called the stable manifold of $S$ and $W^u(S)$ is called the unstable manifold of $S$. Usually, these invariant manifold have an infinite extent and they grow folds and tendrils in the large. The segments of $W^s(S)$ and $W^u(S)$ connected directly to the invariant subset $S$ and short enough to avoid the inclusion of the folds and tendrils will be called the local segments of these invariant manifolds.
  \item {\bf{Homoclinic and heteroclinic intersections:}}
  Assume the situation where we have a phase space $P$ (we think here in particular of the domain of a Poincar\'e map), an invariant subset $S$ and its stable and unstable manifolds $W^s(S)$ and $W^u(S)$. In general systems $W^s(S)$ and $W^u(S)$ only intersect transversally. This means that these two sub-manifold only intersect in the lowest possible dimension and that in any intersection point the tangential spaces of $W^s(S)$ and $W^u(S)$ span the whole tangential space of $P$. Then these intersection points are transverse homoclinic intersection points. If the stable manifold and the unstable manifold come from different unstable invariant subsets then their intersection points are heteroclinic intersection points.
  \item {\bf{Separatrix:}}
  If we have an integrable system, then the stable manifold of an invariant subset can coincide with the unstable manifold of the same or of some other invariant subset. In this case, we call such common stable and unstable manifolds a separatrix. This situation is particularly familiar in Poincar\'e maps of integrable 2-dof systems. Here the Poincar\'e map acts on a 2-dimensional domain. Unstable invariant subsets of dimension zero are hyperbolic fixed points and their stable and unstable manifolds are 1-dimensional curves. Because of integrability the whole domain of the map is foliated into invariant curves of dimension 1 and then the stable and unstable manifolds of hyperbolic fixed points can either go away to infinity or end in an hyperbolic fixed point (either the same one or another one). If we perturb such a situation by a general perturbation which destroys integrability then the separatrix breaks into stable and unstable manifolds with transverse intersections. However, when the perturbation is small then the angle of intersection is small, the forming homoclinic/heteroclinic tangle is very narrow and numerically it is almost impossible to distinguish the created fine chaos strip from the separatrix of the unperturbed case. With increasing perturbation the intersection angle of the broken separatrix increases, the width of the corresponding tangle increases too and gives rise to an increasing chaos strip. Usually in 2-dimensional Poincar\'e maps the large chaos regions grow out of separatrices when we start with an integrable system and let the perturbation grow to large values. This also holds for the restricted map on the NHIM studied in the present article, since this restricted map is a 2-dimensional Poincar\'e map.
  \item {\bf{Index of a saddle:}}
  Imagine a real valued function $F$ defined on a $N$-dimensional real manifold with $N$ coordinates $x_j, j = 1, ..., N $. Let us assume that there is an extremal point of the function and without loss of generality we can assume that the coordinates are chosen such that this extremal point sits in the origin of the coordinate system used, i.e. the extremal point sits at $x_j = 0, j = 1, ..., N$. Now make a power series expansion of the function around the origin and truncate it at second order. Then the quadratic approximation of the function around the extremal point is given in the form
  \begin{equation}
  F(x) = F_0 + \sum_{j=1}^N a_j x_j^2 / 2.
  \end{equation}
  The extremal point is non-degenerate if all $a_j$ are different from zero. And then the index of the extremal point is the number of negative expansion coefficients $a_j$. In our present article the effective potential plays the role of the function $F$.
  \item {\bf{Codimension:}}
  Imagine some sub-manifold $S$ of dimension $k$ embedded in a manifold $M$ of dimension $N$. Then the codimension of $S$ in $M$ is $codim(S) = N - k $.
  \item {\bf{Lyapunov orbits:}}
  Assume an index-$k$ saddle of the effective potential $\Phi$ of a Hamiltonian $N$-dof system. Look at the dynamics generated by the quadratic approximation of the Hamiltonian around this saddle (compare with what we do in the beginning of section \ref{nhims}). The corresponding equations of motion are linear and we can describe the general motion in terms of normal modes obtained after a diagonalization of the equations of motion by an appropriate rotation of the coordinate system. We find $k$ unstable modes leading to motion running away exponentially and $N-k$ stable modes leading to oscillatory motion. If only a single stable mode is excited, then the corresponding motion is a periodic orbit oscillating in this single mode. This periodic orbit is the Lyapunov orbit belonging to this stable mode of motion. They are the skeleton orbits for the whole dynamics in the neighbourhood of the saddle (compare the explanations in subsection \ref{quad}).
  \item {\bf{The monodromy matrix of a periodic orbit:}}
  Assume a periodic orbit $\Gamma$ and an initial condition very close to this periodic orbit, i.e. we study the time-evolution of small deviations from the periodic orbit. Use a coordinate system where some reference point on the periodic orbit is the origin. In these coordinates the general nearby initial point is given by a variational vector $\vec{v}$. To transport the deviation we use the linear variational equations belonging to the equations of motion. After one complete revolution around the periodic orbit $\Gamma$ the resulting transported deviation vector is $\vec{w}$. Because of the linearity of the variational equations there is a matrix $M$ such that $\vec{w} = M \vec{v}$. This matrix $M$ is a property of the periodic orbit $\Gamma$ only and it characterises the stability properties of this periodic orbit. If the dynamics is Hamiltonian, then the matrix $M$ is symplectic. The eigenvalues of $M$ are the eigenvalues of the orbit $\Gamma$ and the eigenvalues of the corresponding fixed point in the Poincar\'e map. If $\lambda$ is a real eigenvalue then also $1/\lambda$ is an eigenvalue and if $\lambda \ne \pm 1$ then the dynamics is unstable (hyperbolic)in the eigenplane belonging to these two eigenvalues $\lambda$ and $1/ \lambda$. If $\lambda$ is a complex eigenvalue of modulus 1, then also the complex conjugate $\bar{\lambda} = 1 / \lambda$ is an eigenvalue and if they are different from $\pm 1$ then the dynamics in the corresponding eigenplane is stable (elliptic). The cases of eigenvalues $\pm 1$ are the parabolic limit cases where bifurcations occur. In principle, there is also the possibility of general complex eigenvalues. However, in our present example of the barred galaxy this case does not occur, therefore we do not explain this case. A very common method to display the various cases in a unified way is to plot the sum of the two related eigenvalues, i.e. to plot $\rm Tr = \lambda + 1/ \lambda$. This is exactly the trace of the corresponding $2 \times 2$ block of the matrix $M$ which belongs to the eigenplane of these two eigenvalues. Also in the elliptic case this trace is real and lies between $-2$ and $+2$. In the unstable case it is either larger than $+2$ for the normal hyperbolic case or it is smaller than $-2$ for the inverse hyperbolic case. The parabolic limit cases have traces $\pm 2$.
\end{itemize}

\bsp
\label{lastpage}

\end{document}